\documentclass[showpacs,amssymb,floatfix,pre,aps,notitlepage]{revtex4-1}

\usepackage[utf8]{inputenc}
\usepackage[english]{babel}

\usepackage{amsmath}
\usepackage{amssymb}
\usepackage{mathtools}

\usepackage{verbatim}

\usepackage{pst-all}
\usepackage{pstricks}
\usepackage[dvips]{graphicx}
\usepackage{psfrag}
\usepackage{subfig}
\usepackage{float}

\usepackage{hyperref}
\usepackage[capitalize]{cleveref}
\usepackage{color}

\newcommand{\suchthat}{\mathrel{\ooalign{$\ni$\cr\kern-1pt$-$\kern-6.5pt$-$}}}
\renewcommand{\r}{\mathbf{r}}
\newcommand{\floor}[1]{\left\lfloor #1 \right\rfloor}
\newcommand{\ceiling}[1]{\left\lceil #1 \right\rceil}

\def\({\left(}
\def\){\right)}
\def\[{\left[}
\def\]{\right]}
\def\|l{\left|}
\def\|{\right|}
\def\<{\left <}
\def\>{\right>}

\newcommand{\Rpar}[1]{\left(#1\right)}
\newcommand{\Bpar}[1]{\left[#1\right]}
\newcommand{\Vpar}[1]{\left\vert #1\right\vert}

\newcommand{\Kpar}[1]{\left\{#1\right\}}

\def\emp{\begin{equation}}
\def\emps{\begin{equation*}}
\def\fin{\end{equation}} 
\def\fins{\end{equation*}}
\def\eemp{\begin{equation}\begin{aligned}} 
\def\eemps{\begin{equation*}\begin{aligned}}
\def\ffin{\end{aligned}\end{equation}} 
\def\ffins{\end{aligned}\end{equation*}}

\newcommand{\defeq}{\mathrel{\mathop:}=}

\newcommand{\ImTc}[5]{\begin{figure}[h!]
\centering
\includegraphics[width=#1\linewidth]{#2.eps}\\
\caption[#5]{\small #4}
\label{fig:#3}
\end{figure}}

\newcommand{\ImT}[4]{\begin{figure}[htp!]
\centering
\includegraphics[width=#1\linewidth]{#2.eps}\\
\caption{\small #4}
\label{fig:#3}
\end{figure}}

\newcommand{\ImTR}[5]{\begin{figure}[htp!]
\centering
\includegraphics[width=#1\linewidth,angle=#5]{#2.eps}\\
\caption{\small #4}
\label{fig:#3}
\end{figure}}

\definecolor{orange}{rgb}{1,0.5,0}
\definecolor{OliveGreen}{rgb}{0,0.6,0}
\definecolor{lightblue}{rgb}{.90,.95,1}
\definecolor{darkblue}{rgb}{0,0,0.5}

\definecolor{airforceblue}{rgb}{0.36, 0.54, 0.66}
\definecolor{aliceblue}{rgb}{0.94, 0.97, 1.0}
\definecolor{alizarin}{rgb}{0.82, 0.1, 0.26}
\definecolor{almond}{rgb}{0.94, 0.87, 0.8}
\definecolor{amaranth}{rgb}{0.9, 0.17, 0.31}
\definecolor{amber}{rgb}{1.0, 0.75, 0.0}
\definecolor{amber(sae/ece)}{rgb}{1.0, 0.49, 0.0}
\definecolor{americanrose}{rgb}{1.0, 0.01, 0.24}
\definecolor{amethyst}{rgb}{0.6, 0.4, 0.8}
\definecolor{anti-flashwhite}{rgb}{0.95, 0.95, 0.96}
\definecolor{antiquebrass}{rgb}{0.8, 0.58, 0.46}
\definecolor{antiquefuchsia}{rgb}{0.57, 0.36, 0.51}
\definecolor{antiquewhite}{rgb}{0.98, 0.92, 0.84}
\definecolor{ao}{rgb}{0.0, 0.0, 1.0}
\definecolor{ao(english)}{rgb}{0.0, 0.5, 0.0}
\definecolor{applegreen}{rgb}{0.55, 0.71, 0.0}
\definecolor{apricot}{rgb}{0.98, 0.81, 0.69}
\definecolor{aqua}{rgb}{0.0, 1.0, 1.0}
\definecolor{aquamarine}{rgb}{0.5, 1.0, 0.83}
\definecolor{armygreen}{rgb}{0.29, 0.33, 0.13}
\definecolor{arsenic}{rgb}{0.23, 0.27, 0.29}
\definecolor{arylideyellow}{rgb}{0.91, 0.84, 0.42}
\definecolor{ashgrey}{rgb}{0.7, 0.75, 0.71}
\definecolor{asparagus}{rgb}{0.53, 0.66, 0.42}
\definecolor{atomictangerine}{rgb}{1.0, 0.6, 0.4}
\definecolor{auburn}{rgb}{0.43, 0.21, 0.1}
\definecolor{aureolin}{rgb}{0.99, 0.93, 0.0}
\definecolor{aurometalsaurus}{rgb}{0.43, 0.5, 0.5}
\definecolor{awesome}{rgb}{1.0, 0.13, 0.32}
\definecolor{azure(colorwheel)}{rgb}{0.0, 0.5, 1.0}
\definecolor{azure(web)(azuremist)}{rgb}{0.94, 1.0, 1.0}
\definecolor{babyblue}{rgb}{0.54, 0.81, 0.94}
\definecolor{babyblueeyes}{rgb}{0.63, 0.79, 0.95}
\definecolor{babypink}{rgb}{0.96, 0.76, 0.76}
\definecolor{ballblue}{rgb}{0.13, 0.67, 0.8}
\definecolor{bananamania}{rgb}{0.98, 0.91, 0.71}
\definecolor{bananayellow}{rgb}{1.0, 0.88, 0.21}
\definecolor{battleshipgrey}{rgb}{0.52, 0.52, 0.51}
\definecolor{bazaar}{rgb}{0.6, 0.47, 0.48}
\definecolor{beaublue}{rgb}{0.74, 0.83, 0.9}
\definecolor{beaver}{rgb}{0.62, 0.51, 0.44}
\definecolor{beige}{rgb}{0.96, 0.96, 0.86}
\definecolor{bisque}{rgb}{1.0, 0.89, 0.77}
\definecolor{bistre}{rgb}{0.24, 0.17, 0.12}
\definecolor{bittersweet}{rgb}{1.0, 0.44, 0.37}
\definecolor{black}{rgb}{0.0, 0.0, 0.0}
\definecolor{blanchedalmond}{rgb}{1.0, 0.92, 0.8}
\definecolor{bleudefrance}{rgb}{0.19, 0.55, 0.91}
\definecolor{blizzardblue}{rgb}{0.67, 0.9, 0.93}
\definecolor{blond}{rgb}{0.98, 0.94, 0.75}
\definecolor{blue}{rgb}{0.0, 0.0, 1.0}
\definecolor{blue(munsell)}{rgb}{0.0, 0.5, 0.69}
\definecolor{blue(ncs)}{rgb}{0.0, 0.53, 0.74}
\definecolor{blue(pigment)}{rgb}{0.2, 0.2, 0.6}
\definecolor{blue(ryb)}{rgb}{0.01, 0.28, 1.0}
\definecolor{bluebell}{rgb}{0.64, 0.64, 0.82}
\definecolor{bluegray}{rgb}{0.4, 0.6, 0.8}
\definecolor{blue-green}{rgb}{0.0, 0.87, 0.87}
\definecolor{blue-violet}{rgb}{0.54, 0.17, 0.89}
\definecolor{blush}{rgb}{0.87, 0.36, 0.51}
\definecolor{bole}{rgb}{0.47, 0.27, 0.23}
\definecolor{bondiblue}{rgb}{0.0, 0.58, 0.71}
\definecolor{bostonuniversityred}{rgb}{0.8, 0.0, 0.0}
\definecolor{brandeisblue}{rgb}{0.0, 0.44, 1.0}
\definecolor{brass}{rgb}{0.71, 0.65, 0.26}
\definecolor{brickred}{rgb}{0.8, 0.25, 0.33}
\definecolor{brightcerulean}{rgb}{0.11, 0.67, 0.84}
\definecolor{brightgreen}{rgb}{0.4, 1.0, 0.0}
\definecolor{brightlavender}{rgb}{0.75, 0.58, 0.89}
\definecolor{brightmaroon}{rgb}{0.76, 0.13, 0.28}
\definecolor{brightpink}{rgb}{1.0, 0.0, 0.5}
\definecolor{brightturquoise}{rgb}{0.03, 0.91, 0.87}
\definecolor{brightube}{rgb}{0.82, 0.62, 0.91}
\definecolor{brilliantlavender}{rgb}{0.96, 0.73, 1.0}
\definecolor{brilliantrose}{rgb}{1.0, 0.33, 0.64}
\definecolor{brinkpink}{rgb}{0.98, 0.38, 0.5}
\definecolor{britishracinggreen}{rgb}{0.0, 0.26, 0.15}
\definecolor{bronze}{rgb}{0.8, 0.5, 0.2}
\definecolor{brown(traditional)}{rgb}{0.59, 0.29, 0.0}
\definecolor{brown(web)}{rgb}{0.65, 0.16, 0.16}
\definecolor{bubblegum}{rgb}{0.99, 0.76, 0.8}
\definecolor{bubbles}{rgb}{0.91, 1.0, 1.0}
\definecolor{buff}{rgb}{0.94, 0.86, 0.51}
\definecolor{bulgarianrose}{rgb}{0.28, 0.02, 0.03}
\definecolor{burgundy}{rgb}{0.5, 0.0, 0.13}
\definecolor{burlywood}{rgb}{0.87, 0.72, 0.53}
\definecolor{burntorange}{rgb}{0.8, 0.33, 0.0}
\definecolor{burntsienna}{rgb}{0.91, 0.45, 0.32}
\definecolor{burntumber}{rgb}{0.54, 0.2, 0.14}
\definecolor{byzantine}{rgb}{0.74, 0.2, 0.64}
\definecolor{byzantium}{rgb}{0.44, 0.16, 0.39}
\definecolor{cadet}{rgb}{0.33, 0.41, 0.47}
\definecolor{cadetblue}{rgb}{0.37, 0.62, 0.63}
\definecolor{cadetgrey}{rgb}{0.57, 0.64, 0.69}
\definecolor{cadmiumgreen}{rgb}{0.0, 0.42, 0.24}
\definecolor{cadmiumorange}{rgb}{0.93, 0.53, 0.18}
\definecolor{cadmiumred}{rgb}{0.89, 0.0, 0.13}
\definecolor{cadmiumyellow}{rgb}{1.0, 0.96, 0.0}
\definecolor{calpolypomonagreen}{rgb}{0.12, 0.3, 0.17}
\definecolor{cambridgeblue}{rgb}{0.64, 0.76, 0.68}
\definecolor{camel}{rgb}{0.76, 0.6, 0.42}
\definecolor{camouflagegreen}{rgb}{0.47, 0.53, 0.42}
\definecolor{canaryyellow}{rgb}{1.0, 0.94, 0.0}
\definecolor{candyapplered}{rgb}{1.0, 0.03, 0.0}
\definecolor{candypink}{rgb}{0.89, 0.44, 0.48}
\definecolor{capri}{rgb}{0.0, 0.75, 1.0}
\definecolor{caputmortuum}{rgb}{0.35, 0.15, 0.13}
\definecolor{cardinal}{rgb}{0.77, 0.12, 0.23}
\definecolor{caribbeangreen}{rgb}{0.0, 0.8, 0.6}
\definecolor{carmine}{rgb}{0.59, 0.0, 0.09}
\definecolor{carminepink}{rgb}{0.92, 0.3, 0.26}
\definecolor{carminered}{rgb}{1.0, 0.0, 0.22}
\definecolor{carnationpink}{rgb}{1.0, 0.65, 0.79}
\definecolor{carnelian}{rgb}{0.7, 0.11, 0.11}
\definecolor{carolinablue}{rgb}{0.6, 0.73, 0.89}
\definecolor{carrotorange}{rgb}{0.93, 0.57, 0.13}
\definecolor{ceil}{rgb}{0.57, 0.63, 0.81}
\definecolor{celadon}{rgb}{0.67, 0.88, 0.69}
\definecolor{celestialblue}{rgb}{0.29, 0.59, 0.82}
\definecolor{cerise}{rgb}{0.87, 0.19, 0.39}
\definecolor{cerisepink}{rgb}{0.93, 0.23, 0.51}
\definecolor{cerulean}{rgb}{0.0, 0.48, 0.65}
\definecolor{ceruleanblue}{rgb}{0.16, 0.32, 0.75}
\definecolor{chamoisee}{rgb}{0.63, 0.47, 0.35}
\definecolor{champagne}{rgb}{0.97, 0.91, 0.81}
\definecolor{charcoal}{rgb}{0.21, 0.27, 0.31}
\definecolor{chartreuse(traditional)}{rgb}{0.87, 1.0, 0.0}
\definecolor{chartreuse(web)}{rgb}{0.5, 1.0, 0.0}
\definecolor{cherryblossompink}{rgb}{1.0, 0.72, 0.77}
\definecolor{chestnut}{rgb}{0.8, 0.36, 0.36}
\definecolor{chocolate(traditional)}{rgb}{0.48, 0.25, 0.0}
\definecolor{chocolate(web)}{rgb}{0.82, 0.41, 0.12}
\definecolor{chromeyellow}{rgb}{1.0, 0.65, 0.0}
\definecolor{cinereous}{rgb}{0.6, 0.51, 0.48}
\definecolor{cinnabar}{rgb}{0.89, 0.26, 0.2}
\definecolor{cinnamon}{rgb}{0.82, 0.41, 0.12}
\definecolor{citrine}{rgb}{0.89, 0.82, 0.04}
\definecolor{classicrose}{rgb}{0.98, 0.8, 0.91}
\definecolor{cobalt}{rgb}{0.0, 0.28, 0.67}
\definecolor{cocoabrown}{rgb}{0.82, 0.41, 0.12}
\definecolor{columbiablue}{rgb}{0.61, 0.87, 1.0}
\definecolor{coolblack}{rgb}{0.0, 0.18, 0.39}
\definecolor{coolgrey}{rgb}{0.55, 0.57, 0.67}
\definecolor{copper}{rgb}{0.72, 0.45, 0.2}
\definecolor{copperrose}{rgb}{0.6, 0.4, 0.4}
\definecolor{coquelicot}{rgb}{1.0, 0.22, 0.0}
\definecolor{coral}{rgb}{1.0, 0.5, 0.31}
\definecolor{coralpink}{rgb}{0.97, 0.51, 0.47}
\definecolor{coralred}{rgb}{1.0, 0.25, 0.25}
\definecolor{cordovan}{rgb}{0.54, 0.25, 0.27}
\definecolor{corn}{rgb}{0.98, 0.93, 0.36}
\definecolor{cornellred}{rgb}{0.7, 0.11, 0.11}
\definecolor{cornflowerblue}{rgb}{0.39, 0.58, 0.93}
\definecolor{cornsilk}{rgb}{1.0, 0.97, 0.86}
\definecolor{cosmiclatte}{rgb}{1.0, 0.97, 0.91}
\definecolor{cottoncandy}{rgb}{1.0, 0.74, 0.85}
\definecolor{cream}{rgb}{1.0, 0.99, 0.82}
\definecolor{crimson}{rgb}{0.86, 0.08, 0.24}
\definecolor{crimsonglory}{rgb}{0.75, 0.0, 0.2}
\definecolor{cyan}{rgb}{0.0, 1.0, 1.0}
\definecolor{cyan(process)}{rgb}{0.0, 0.72, 0.92}
\definecolor{daffodil}{rgb}{1.0, 1.0, 0.19}
\definecolor{dandelion}{rgb}{0.94, 0.88, 0.19}
\definecolor{darkblue}{rgb}{0.0, 0.0, 0.55}
\definecolor{darkbrown}{rgb}{0.4, 0.26, 0.13}
\definecolor{darkbyzantium}{rgb}{0.36, 0.22, 0.33}
\definecolor{darkcandyapplered}{rgb}{0.64, 0.0, 0.0}
\definecolor{darkcerulean}{rgb}{0.03, 0.27, 0.49}
\definecolor{darkchampagne}{rgb}{0.76, 0.7, 0.5}
\definecolor{darkchestnut}{rgb}{0.6, 0.41, 0.38}
\definecolor{darkcoral}{rgb}{0.8, 0.36, 0.27}
\definecolor{darkcyan}{rgb}{0.0, 0.55, 0.55}
\definecolor{darkelectricblue}{rgb}{0.33, 0.41, 0.47}
\definecolor{darkgoldenrod}{rgb}{0.72, 0.53, 0.04}
\definecolor{darkgray}{rgb}{0.66, 0.66, 0.66}
\definecolor{darkgreen}{rgb}{0.0, 0.2, 0.13}
\definecolor{darkjunglegreen}{rgb}{0.1, 0.14, 0.13}
\definecolor{darkkhaki}{rgb}{0.74, 0.72, 0.42}
\definecolor{darklava}{rgb}{0.28, 0.24, 0.2}
\definecolor{darklavender}{rgb}{0.45, 0.31, 0.59}
\definecolor{darkmagenta}{rgb}{0.55, 0.0, 0.55}
\definecolor{darkmidnightblue}{rgb}{0.0, 0.2, 0.4}
\definecolor{darkolivegreen}{rgb}{0.33, 0.42, 0.18}
\definecolor{darkorange}{rgb}{1.0, 0.55, 0.0}
\definecolor{darkorchid}{rgb}{0.6, 0.2, 0.8}
\definecolor{darkpastelblue}{rgb}{0.47, 0.62, 0.8}
\definecolor{darkpastelgreen}{rgb}{0.01, 0.75, 0.24}
\definecolor{darkpastelpurple}{rgb}{0.59, 0.44, 0.84}
\definecolor{darkpastelred}{rgb}{0.76, 0.23, 0.13}
\definecolor{darkpink}{rgb}{0.91, 0.33, 0.5}
\definecolor{darkpowderblue}{rgb}{0.0, 0.2, 0.6}
\definecolor{darkraspberry}{rgb}{0.53, 0.15, 0.34}
\definecolor{darkred}{rgb}{0.55, 0.0, 0.0}
\definecolor{darksalmon}{rgb}{0.91, 0.59, 0.48}
\definecolor{darkscarlet}{rgb}{0.34, 0.01, 0.1}
\definecolor{darkseagreen}{rgb}{0.56, 0.74, 0.56}
\definecolor{darksienna}{rgb}{0.24, 0.08, 0.08}
\definecolor{darkslateblue}{rgb}{0.28, 0.24, 0.55}
\definecolor{darkslategray}{rgb}{0.18, 0.31, 0.31}
\definecolor{darkspringgreen}{rgb}{0.09, 0.45, 0.27}
\definecolor{darktan}{rgb}{0.57, 0.51, 0.32}
\definecolor{darktangerine}{rgb}{1.0, 0.66, 0.07}
\definecolor{darktaupe}{rgb}{0.28, 0.24, 0.2}
\definecolor{darkterracotta}{rgb}{0.8, 0.31, 0.36}
\definecolor{darkturquoise}{rgb}{0.0, 0.81, 0.82}
\definecolor{darkviolet}{rgb}{0.58, 0.0, 0.83}
\definecolor{dartmouthgreen}{rgb}{0.05, 0.5, 0.06}
\definecolor{davysgrey}{rgb}{0.33, 0.33, 0.33}
\definecolor{debianred}{rgb}{0.84, 0.04, 0.33}
\definecolor{deepcarmine}{rgb}{0.66, 0.13, 0.24}
\definecolor{deepcarminepink}{rgb}{0.94, 0.19, 0.22}
\definecolor{deepcarrotorange}{rgb}{0.91, 0.41, 0.17}
\definecolor{deepcerise}{rgb}{0.85, 0.2, 0.53}
\definecolor{deepchampagne}{rgb}{0.98, 0.84, 0.65}
\definecolor{deepchestnut}{rgb}{0.73, 0.31, 0.28}
\definecolor{deepfuchsia}{rgb}{0.76, 0.33, 0.76}
\definecolor{deepjunglegreen}{rgb}{0.0, 0.29, 0.29}
\definecolor{deeplilac}{rgb}{0.6, 0.33, 0.73}
\definecolor{deepmagenta}{rgb}{0.8, 0.0, 0.8}
\definecolor{deeppeach}{rgb}{1.0, 0.8, 0.64}
\definecolor{deeppink}{rgb}{1.0, 0.08, 0.58}
\definecolor{deepsaffron}{rgb}{1.0, 0.6, 0.2}
\definecolor{deepskyblue}{rgb}{0.0, 0.75, 1.0}
\definecolor{denim}{rgb}{0.08, 0.38, 0.74}
\definecolor{desert}{rgb}{0.76, 0.6, 0.42}
\definecolor{desertsand}{rgb}{0.93, 0.79, 0.69}
\definecolor{dimgray}{rgb}{0.41, 0.41, 0.41}
\definecolor{dodgerblue}{rgb}{0.12, 0.56, 1.0}
\definecolor{dogwoodrose}{rgb}{0.84, 0.09, 0.41}
\definecolor{dollarbill}{rgb}{0.52, 0.73, 0.4}
\definecolor{drab}{rgb}{0.59, 0.44, 0.09}
\definecolor{dukeblue}{rgb}{0.0, 0.0, 0.61}
\definecolor{earthyellow}{rgb}{0.88, 0.66, 0.37}
\definecolor{ecru}{rgb}{0.76, 0.7, 0.5}
\definecolor{eggplant}{rgb}{0.38, 0.25, 0.32}
\definecolor{eggshell}{rgb}{0.94, 0.92, 0.84}
\definecolor{egyptianblue}{rgb}{0.06, 0.2, 0.65}
\definecolor{electricblue}{rgb}{0.49, 0.98, 1.0}
\definecolor{electriccrimson}{rgb}{1.0, 0.0, 0.25}
\definecolor{electriccyan}{rgb}{0.0, 1.0, 1.0}
\definecolor{electricgreen}{rgb}{0.0, 1.0, 0.0}
\definecolor{electricindigo}{rgb}{0.44, 0.0, 1.0}
\definecolor{electriclavender}{rgb}{0.96, 0.73, 1.0}
\definecolor{electriclime}{rgb}{0.8, 1.0, 0.0}
\definecolor{electricpurple}{rgb}{0.75, 0.0, 1.0}
\definecolor{electricultramarine}{rgb}{0.25, 0.0, 1.0}
\definecolor{electricviolet}{rgb}{0.56, 0.0, 1.0}
\definecolor{electricyellow}{rgb}{1.0, 1.0, 0.0}
\definecolor{emerald}{rgb}{0.31, 0.78, 0.47}
\definecolor{etonblue}{rgb}{0.59, 0.78, 0.64}
\definecolor{fallow}{rgb}{0.76, 0.6, 0.42}
\definecolor{falured}{rgb}{0.5, 0.09, 0.09}
\definecolor{fandango}{rgb}{0.71, 0.2, 0.54}
\definecolor{fashionfuchsia}{rgb}{0.96, 0.0, 0.63}
\definecolor{fawn}{rgb}{0.9, 0.67, 0.44}
\definecolor{feldgrau}{rgb}{0.3, 0.36, 0.33}
\definecolor{ferngreen}{rgb}{0.31, 0.47, 0.26}
\definecolor{ferrarired}{rgb}{1.0, 0.11, 0.0}
\definecolor{fielddrab}{rgb}{0.42, 0.33, 0.12}
\definecolor{firebrick}{rgb}{0.7, 0.13, 0.13}
\definecolor{fireenginered}{rgb}{0.81, 0.09, 0.13}
\definecolor{flame}{rgb}{0.89, 0.35, 0.13}
\definecolor{flamingopink}{rgb}{0.99, 0.56, 0.67}
\definecolor{flavescent}{rgb}{0.97, 0.91, 0.56}
\definecolor{flax}{rgb}{0.93, 0.86, 0.51}
\definecolor{floralwhite}{rgb}{1.0, 0.98, 0.94}
\definecolor{fluorescentorange}{rgb}{1.0, 0.75, 0.0}
\definecolor{fluorescentpink}{rgb}{1.0, 0.08, 0.58}
\definecolor{fluorescentyellow}{rgb}{0.8, 1.0, 0.0}
\definecolor{folly}{rgb}{1.0, 0.0, 0.31}
\definecolor{forestgreen(traditional)}{rgb}{0.0, 0.27, 0.13}
\definecolor{forestgreen(web)}{rgb}{0.13, 0.55, 0.13}
\definecolor{frenchbeige}{rgb}{0.65, 0.48, 0.36}
\definecolor{frenchblue}{rgb}{0.0, 0.45, 0.73}
\definecolor{frenchlilac}{rgb}{0.53, 0.38, 0.56}
\definecolor{frenchrose}{rgb}{0.96, 0.29, 0.54}
\definecolor{fuchsia}{rgb}{1.0, 0.0, 1.0}
\definecolor{fuchsiapink}{rgb}{1.0, 0.47, 1.0}
\definecolor{fulvous}{rgb}{0.86, 0.52, 0.0}
\definecolor{fuzzywuzzy}{rgb}{0.8, 0.4, 0.4}
\definecolor{gainsboro}{rgb}{0.86, 0.86, 0.86}
\definecolor{gamboge}{rgb}{0.89, 0.61, 0.06}
\definecolor{ghostwhite}{rgb}{0.97, 0.97, 1.0}
\definecolor{ginger}{rgb}{0.69, 0.4, 0.0}
\definecolor{glaucous}{rgb}{0.38, 0.51, 0.71}
\definecolor{gold(metallic)}{rgb}{0.83, 0.69, 0.22}
\definecolor{gold(web)(golden)}{rgb}{1.0, 0.84, 0.0}
\definecolor{goldenbrown}{rgb}{0.6, 0.4, 0.08}
\definecolor{goldenpoppy}{rgb}{0.99, 0.76, 0.0}
\definecolor{goldenyellow}{rgb}{1.0, 0.87, 0.0}
\definecolor{goldenrod}{rgb}{0.85, 0.65, 0.13}
\definecolor{grannysmithapple}{rgb}{0.66, 0.89, 0.63}
\definecolor{gray}{rgb}{0.5, 0.5, 0.5}
\definecolor{gray(html/cssgray)}{rgb}{0.5, 0.5, 0.5}
\definecolor{gray(x11gray)}{rgb}{0.75, 0.75, 0.75}
\definecolor{gray-asparagus}{rgb}{0.27, 0.35, 0.27}
\definecolor{green(colorwheel)(x11green)}{rgb}{0.0, 1.0, 0.0}
\definecolor{green(html/cssgreen)}{rgb}{0.0, 0.5, 0.0}
\definecolor{green(munsell)}{rgb}{0.0, 0.66, 0.47}
\definecolor{green(ncs)}{rgb}{0.0, 0.62, 0.42}
\definecolor{green(pigment)}{rgb}{0.0, 0.65, 0.31}
\definecolor{green(ryb)}{rgb}{0.4, 0.69, 0.2}
\definecolor{green-yellow}{rgb}{0.68, 1.0, 0.18}
\definecolor{grullo}{rgb}{0.66, 0.6, 0.53}
\definecolor{guppiegreen}{rgb}{0.0, 1.0, 0.5}
\definecolor{halayaube}{rgb}{0.4, 0.22, 0.33}
\definecolor{hanblue}{rgb}{0.27, 0.42, 0.81}
\definecolor{hanpurple}{rgb}{0.32, 0.09, 0.98}
\definecolor{hansayellow}{rgb}{0.91, 0.84, 0.42}
\definecolor{harlequin}{rgb}{0.25, 1.0, 0.0}
\definecolor{harvardcrimson}{rgb}{0.79, 0.0, 0.09}
\definecolor{harvestgold}{rgb}{0.85, 0.57, 0.0}
\definecolor{heartgold}{rgb}{0.5, 0.5, 0.0}
\definecolor{heliotrope}{rgb}{0.87, 0.45, 1.0}
\definecolor{hollywoodcerise}{rgb}{0.96, 0.0, 0.63}
\definecolor{honeydew}{rgb}{0.94, 1.0, 0.94}
\definecolor{hookersgreen}{rgb}{0.0, 0.44, 0.0}
\definecolor{hotmagenta}{rgb}{1.0, 0.11, 0.81}
\definecolor{hotpink}{rgb}{1.0, 0.41, 0.71}
\definecolor{huntergreen}{rgb}{0.21, 0.37, 0.23}
\definecolor{iceberg}{rgb}{0.44, 0.65, 0.82}
\definecolor{icterine}{rgb}{0.99, 0.97, 0.37}
\definecolor{inchworm}{rgb}{0.7, 0.93, 0.36}
\definecolor{indiagreen}{rgb}{0.07, 0.53, 0.03}
\definecolor{indianred}{rgb}{0.8, 0.36, 0.36}
\definecolor{indianyellow}{rgb}{0.89, 0.66, 0.34}
\definecolor{indigo(dye)}{rgb}{0.0, 0.25, 0.42}
\definecolor{indigo(web)}{rgb}{0.29, 0.0, 0.51}
\definecolor{internationalkleinblue}{rgb}{0.0, 0.18, 0.65}
\definecolor{internationalorange}{rgb}{1.0, 0.31, 0.0}
\definecolor{iris}{rgb}{0.35, 0.31, 0.81}
\definecolor{isabelline}{rgb}{0.96, 0.94, 0.93}
\definecolor{islamicgreen}{rgb}{0.0, 0.56, 0.0}
\definecolor{ivory}{rgb}{1.0, 1.0, 0.94}
\definecolor{jade}{rgb}{0.0, 0.66, 0.42}
\definecolor{jasper}{rgb}{0.84, 0.23, 0.24}
\definecolor{jazzberryjam}{rgb}{0.65, 0.04, 0.37}
\definecolor{jonquil}{rgb}{0.98, 0.85, 0.37}
\definecolor{junebud}{rgb}{0.74, 0.85, 0.34}
\definecolor{junglegreen}{rgb}{0.16, 0.67, 0.53}
\definecolor{kellygreen}{rgb}{0.3, 0.73, 0.09}
\definecolor{khaki(html/css)(khaki)}{rgb}{0.76, 0.69, 0.57}
\definecolor{khaki(x11)(lightkhaki)}{rgb}{0.94, 0.9, 0.55}
\definecolor{lasallegreen}{rgb}{0.03, 0.47, 0.19}
\definecolor{languidlavender}{rgb}{0.84, 0.79, 0.87}
\definecolor{lapislazuli}{rgb}{0.15, 0.38, 0.61}
\definecolor{laserlemon}{rgb}{1.0, 1.0, 0.13}
\definecolor{lava}{rgb}{0.81, 0.06, 0.13}
\definecolor{lavender(floral)}{rgb}{0.71, 0.49, 0.86}
\definecolor{lavender(web)}{rgb}{0.9, 0.9, 0.98}
\definecolor{lavenderblue}{rgb}{0.8, 0.8, 1.0}
\definecolor{lavenderblush}{rgb}{1.0, 0.94, 0.96}
\definecolor{lavendergray}{rgb}{0.77, 0.76, 0.82}
\definecolor{lavenderindigo}{rgb}{0.58, 0.34, 0.92}
\definecolor{lavendermagenta}{rgb}{0.93, 0.51, 0.93}
\definecolor{lavendermist}{rgb}{0.9, 0.9, 0.98}
\definecolor{lavenderpink}{rgb}{0.98, 0.68, 0.82}
\definecolor{lavenderpurple}{rgb}{0.59, 0.48, 0.71}
\definecolor{lavenderrose}{rgb}{0.98, 0.63, 0.89}
\definecolor{lawngreen}{rgb}{0.49, 0.99, 0.0}
\definecolor{lemon}{rgb}{1.0, 0.97, 0.0}
\definecolor{lemonchiffon}{rgb}{1.0, 0.98, 0.8}
\definecolor{lightapricot}{rgb}{0.99, 0.84, 0.69}
\definecolor{lightblue}{rgb}{0.68, 0.85, 0.9}
\definecolor{lightbrown}{rgb}{0.71, 0.4, 0.11}
\definecolor{lightcarminepink}{rgb}{0.9, 0.4, 0.38}
\definecolor{lightcoral}{rgb}{0.94, 0.5, 0.5}
\definecolor{lightcornflowerblue}{rgb}{0.6, 0.81, 0.93}
\definecolor{lightcyan}{rgb}{0.88, 1.0, 1.0}
\definecolor{lightfuchsiapink}{rgb}{0.98, 0.52, 0.9}
\definecolor{lightgoldenrodyellow}{rgb}{0.98, 0.98, 0.82}
\definecolor{lightgray}{rgb}{0.83, 0.83, 0.83}
\definecolor{lightgreen}{rgb}{0.56, 0.93, 0.56}
\definecolor{lightkhaki}{rgb}{0.94, 0.9, 0.55}
\definecolor{lightmauve}{rgb}{0.86, 0.82, 1.0}
\definecolor{lightpastelpurple}{rgb}{0.69, 0.61, 0.85}
\definecolor{lightpink}{rgb}{1.0, 0.71, 0.76}
\definecolor{lightsalmon}{rgb}{1.0, 0.63, 0.48}
\definecolor{lightsalmonpink}{rgb}{1.0, 0.6, 0.6}
\definecolor{lightseagreen}{rgb}{0.13, 0.7, 0.67}
\definecolor{lightskyblue}{rgb}{0.53, 0.81, 0.98}
\definecolor{lightslategray}{rgb}{0.47, 0.53, 0.6}
\definecolor{lighttaupe}{rgb}{0.7, 0.55, 0.43}
\definecolor{lightthulianpink}{rgb}{0.9, 0.56, 0.67}
\definecolor{lightyellow}{rgb}{1.0, 1.0, 0.88}
\definecolor{lilac}{rgb}{0.78, 0.64, 0.78}
\definecolor{lime(colorwheel)}{rgb}{0.75, 1.0, 0.0}
\definecolor{lime(web)(x11green)}{rgb}{0.0, 1.0, 0.0}
\definecolor{limegreen}{rgb}{0.2, 0.8, 0.2}
\definecolor{lincolngreen}{rgb}{0.11, 0.35, 0.02}
\definecolor{linen}{rgb}{0.98, 0.94, 0.9}
\definecolor{liver}{rgb}{0.33, 0.29, 0.31}
\definecolor{lust}{rgb}{0.9, 0.13, 0.13}
\definecolor{macaroniandcheese}{rgb}{1.0, 0.74, 0.53}
\definecolor{magenta}{rgb}{1.0, 0.0, 1.0}
\definecolor{magenta(dye)}{rgb}{0.79, 0.08, 0.48}
\definecolor{magenta(process)}{rgb}{1.0, 0.0, 0.56}
\definecolor{magicmint}{rgb}{0.67, 0.94, 0.82}
\definecolor{magnolia}{rgb}{0.97, 0.96, 1.0}
\definecolor{mahogany}{rgb}{0.75, 0.25, 0.0}
\definecolor{maize}{rgb}{0.98, 0.93, 0.37}
\definecolor{majorelleblue}{rgb}{0.38, 0.31, 0.86}
\definecolor{malachite}{rgb}{0.04, 0.85, 0.32}
\definecolor{manatee}{rgb}{0.59, 0.6, 0.67}
\definecolor{mangotango}{rgb}{1.0, 0.51, 0.26}
\definecolor{maroon(html/css)}{rgb}{0.5, 0.0, 0.0}
\definecolor{maroon(x11)}{rgb}{0.69, 0.19, 0.38}
\definecolor{mauve}{rgb}{0.88, 0.69, 1.0}
\definecolor{mauvetaupe}{rgb}{0.57, 0.37, 0.43}
\definecolor{mauvelous}{rgb}{0.94, 0.6, 0.67}
\definecolor{mayablue}{rgb}{0.45, 0.76, 0.98}
\definecolor{meatbrown}{rgb}{0.9, 0.72, 0.23}
\definecolor{mediumaquamarine}{rgb}{0.4, 0.8, 0.67}
\definecolor{mediumblue}{rgb}{0.0, 0.0, 0.8}
\definecolor{mediumcandyapplered}{rgb}{0.89, 0.02, 0.17}
\definecolor{mediumcarmine}{rgb}{0.69, 0.25, 0.21}
\definecolor{mediumchampagne}{rgb}{0.95, 0.9, 0.67}
\definecolor{mediumelectricblue}{rgb}{0.01, 0.31, 0.59}
\definecolor{mediumjunglegreen}{rgb}{0.11, 0.21, 0.18}
\definecolor{mediumlavendermagenta}{rgb}{0.8, 0.6, 0.8}
\definecolor{mediumorchid}{rgb}{0.73, 0.33, 0.83}
\definecolor{mediumpersianblue}{rgb}{0.0, 0.4, 0.65}
\definecolor{mediumpurple}{rgb}{0.58, 0.44, 0.86}
\definecolor{mediumred-violet}{rgb}{0.73, 0.2, 0.52}
\definecolor{mediumseagreen}{rgb}{0.24, 0.7, 0.44}
\definecolor{mediumslateblue}{rgb}{0.48, 0.41, 0.93}
\definecolor{mediumspringbud}{rgb}{0.79, 0.86, 0.54}
\definecolor{mediumspringgreen}{rgb}{0.0, 0.98, 0.6}
\definecolor{mediumtaupe}{rgb}{0.4, 0.3, 0.28}
\definecolor{mediumtealblue}{rgb}{0.0, 0.33, 0.71}
\definecolor{mediumturquoise}{rgb}{0.28, 0.82, 0.8}
\definecolor{mediumviolet-red}{rgb}{0.78, 0.08, 0.52}
\definecolor{melon}{rgb}{0.99, 0.74, 0.71}
\definecolor{midnightblue}{rgb}{0.1, 0.1, 0.44}
\definecolor{midnightgreen(eaglegreen)}{rgb}{0.0, 0.29, 0.33}
\definecolor{mikadoyellow}{rgb}{1.0, 0.77, 0.05}
\definecolor{mint}{rgb}{0.24, 0.71, 0.54}
\definecolor{mintcream}{rgb}{0.96, 1.0, 0.98}
\definecolor{mintgreen}{rgb}{0.6, 1.0, 0.6}
\definecolor{mistyrose}{rgb}{1.0, 0.89, 0.88}
\definecolor{moccasin}{rgb}{0.98, 0.92, 0.84}
\definecolor{modebeige}{rgb}{0.59, 0.44, 0.09}
\definecolor{moonstoneblue}{rgb}{0.45, 0.66, 0.76}
\definecolor{mordantred19}{rgb}{0.68, 0.05, 0.0}
\definecolor{mossgreen}{rgb}{0.68, 0.87, 0.68}
\definecolor{mountainmeadow}{rgb}{0.19, 0.73, 0.56}
\definecolor{mountbattenpink}{rgb}{0.6, 0.48, 0.55}
\definecolor{mulberry}{rgb}{0.77, 0.29, 0.55}
\definecolor{mustard}{rgb}{1.0, 0.86, 0.35}
\definecolor{myrtle}{rgb}{0.13, 0.26, 0.12}
\definecolor{msugreen}{rgb}{0.09, 0.27, 0.23}
\definecolor{nadeshikopink}{rgb}{0.96, 0.68, 0.78}
\definecolor{napiergreen}{rgb}{0.16, 0.5, 0.0}
\definecolor{naplesyellow}{rgb}{0.98, 0.85, 0.37}
\definecolor{navajowhite}{rgb}{1.0, 0.87, 0.68}
\definecolor{navyblue}{rgb}{0.0, 0.0, 0.5}
\definecolor{neoncarrot}{rgb}{1.0, 0.64, 0.26}
\definecolor{neonfuchsia}{rgb}{1.0, 0.25, 0.39}
\definecolor{neongreen}{rgb}{0.22, 0.88, 0.08}
\definecolor{non-photoblue}{rgb}{0.64, 0.87, 0.93}
\definecolor{oceanboatblue}{rgb}{0.0, 0.47, 0.75}
\definecolor{ochre}{rgb}{0.8, 0.47, 0.13}
\definecolor{officegreen}{rgb}{0.0, 0.5, 0.0}
\definecolor{oldgold}{rgb}{0.81, 0.71, 0.23}
\definecolor{oldlace}{rgb}{0.99, 0.96, 0.9}
\definecolor{oldlavender}{rgb}{0.47, 0.41, 0.47}
\definecolor{oldmauve}{rgb}{0.4, 0.19, 0.28}
\definecolor{oldrose}{rgb}{0.75, 0.5, 0.51}
\definecolor{olive}{rgb}{0.5, 0.5, 0.0}
\definecolor{olivedrabN3}{rgb}{0.42, 0.56, 0.14}
\definecolor{olivedrabN7}{rgb}{0.24, 0.2, 0.12}
\definecolor{olivine}{rgb}{0.6, 0.73, 0.45}
\definecolor{onyx}{rgb}{0.06, 0.06, 0.06}
\definecolor{operamauve}{rgb}{0.72, 0.52, 0.65}
\definecolor{orange(colorwheel)}{rgb}{1.0, 0.5, 0.0}
\definecolor{orange(ryb)}{rgb}{0.98, 0.6, 0.01}
\definecolor{orange(webcolor)}{rgb}{1.0, 0.65, 0.0}
\definecolor{orangepeel}{rgb}{1.0, 0.62, 0.0}
\definecolor{orange-red}{rgb}{1.0, 0.27, 0.0}
\definecolor{orchid}{rgb}{0.85, 0.44, 0.84}
\definecolor{otterbrown}{rgb}{0.4, 0.26, 0.13}
\definecolor{outerspace}{rgb}{0.25, 0.29, 0.3}
\definecolor{outrageousorange}{rgb}{1.0, 0.43, 0.29}
\definecolor{oxfordblue}{rgb}{0.0, 0.13, 0.28}
\definecolor{oucrimsonred}{rgb}{0.6, 0.0, 0.0}
\definecolor{pakistangreen}{rgb}{0.0, 0.4, 0.0}
\definecolor{palatinateblue}{rgb}{0.15, 0.23, 0.89}
\definecolor{palatinatepurple}{rgb}{0.41, 0.16, 0.38}
\definecolor{paleaqua}{rgb}{0.74, 0.83, 0.9}
\definecolor{paleblue}{rgb}{0.69, 0.93, 0.93}
\definecolor{palebrown}{rgb}{0.6, 0.46, 0.33}
\definecolor{palecarmine}{rgb}{0.69, 0.25, 0.21}
\definecolor{palecerulean}{rgb}{0.61, 0.77, 0.89}
\definecolor{palechestnut}{rgb}{0.87, 0.68, 0.69}
\definecolor{palecopper}{rgb}{0.85, 0.54, 0.4}
\definecolor{palecornflowerblue}{rgb}{0.67, 0.8, 0.94}
\definecolor{palegold}{rgb}{0.9, 0.75, 0.54}
\definecolor{palegoldenrod}{rgb}{0.93, 0.91, 0.67}
\definecolor{palegreen}{rgb}{0.6, 0.98, 0.6}
\definecolor{palemagenta}{rgb}{0.98, 0.52, 0.9}
\definecolor{palepink}{rgb}{0.98, 0.85, 0.87}
\definecolor{paleplum}{rgb}{0.8, 0.6, 0.8}
\definecolor{palered-violet}{rgb}{0.86, 0.44, 0.58}
\definecolor{palerobineggblue}{rgb}{0.59, 0.87, 0.82}
\definecolor{palesilver}{rgb}{0.79, 0.75, 0.73}
\definecolor{palespringbud}{rgb}{0.93, 0.92, 0.74}
\definecolor{paletaupe}{rgb}{0.74, 0.6, 0.49}
\definecolor{paleviolet-red}{rgb}{0.86, 0.44, 0.58}
\definecolor{pansypurple}{rgb}{0.47, 0.09, 0.29}
\definecolor{papayawhip}{rgb}{1.0, 0.94, 0.84}
\definecolor{parisgreen}{rgb}{0.31, 0.78, 0.47}
\definecolor{pastelblue}{rgb}{0.68, 0.78, 0.81}
\definecolor{pastelbrown}{rgb}{0.51, 0.41, 0.33}
\definecolor{pastelgray}{rgb}{0.81, 0.81, 0.77}
\definecolor{pastelgreen}{rgb}{0.47, 0.87, 0.47}
\definecolor{pastelmagenta}{rgb}{0.96, 0.6, 0.76}
\definecolor{pastelorange}{rgb}{1.0, 0.7, 0.28}
\definecolor{pastelpink}{rgb}{1.0, 0.82, 0.86}
\definecolor{pastelpurple}{rgb}{0.7, 0.62, 0.71}
\definecolor{pastelred}{rgb}{1.0, 0.41, 0.38}
\definecolor{pastelviolet}{rgb}{0.8, 0.6, 0.79}
\definecolor{pastelyellow}{rgb}{0.99, 0.99, 0.59}
\definecolor{patriarch}{rgb}{0.5, 0.0, 0.5}
\definecolor{paynesgrey}{rgb}{0.25, 0.25, 0.28}
\definecolor{peach}{rgb}{1.0, 0.9, 0.71}
\definecolor{peach-orange}{rgb}{1.0, 0.8, 0.6}
\definecolor{peachpuff}{rgb}{1.0, 0.85, 0.73}
\definecolor{peach-yellow}{rgb}{0.98, 0.87, 0.68}
\definecolor{pear}{rgb}{0.82, 0.89, 0.19}
\definecolor{pearl}{rgb}{0.94, 0.92, 0.84}
\definecolor{peridot}{rgb}{0.9, 0.89, 0.0}
\definecolor{periwinkle}{rgb}{0.8, 0.8, 1.0}
\definecolor{persianblue}{rgb}{0.11, 0.22, 0.73}
\definecolor{persiangreen}{rgb}{0.0, 0.65, 0.58}
\definecolor{persianindigo}{rgb}{0.2, 0.07, 0.48}
\definecolor{persianorange}{rgb}{0.85, 0.56, 0.35}
\definecolor{peru}{rgb}{0.8, 0.52, 0.25}
\definecolor{persianpink}{rgb}{0.97, 0.5, 0.75}
\definecolor{persianplum}{rgb}{0.44, 0.11, 0.11}
\definecolor{persianred}{rgb}{0.8, 0.2, 0.2}
\definecolor{persianrose}{rgb}{1.0, 0.16, 0.64}
\definecolor{persimmon}{rgb}{0.93, 0.35, 0.0}
\definecolor{phlox}{rgb}{0.87, 0.0, 1.0}
\definecolor{phthaloblue}{rgb}{0.0, 0.06, 0.54}
\definecolor{phthalogreen}{rgb}{0.07, 0.21, 0.14}
\definecolor{piggypink}{rgb}{0.99, 0.87, 0.9}
\definecolor{pinegreen}{rgb}{0.0, 0.47, 0.44}
\definecolor{pink}{rgb}{1.0, 0.75, 0.8}
\definecolor{pink-orange}{rgb}{1.0, 0.6, 0.4}
\definecolor{pinkpearl}{rgb}{0.91, 0.67, 0.81}
\definecolor{pinksherbet}{rgb}{0.97, 0.56, 0.65}
\definecolor{pistachio}{rgb}{0.58, 0.77, 0.45}
\definecolor{platinum}{rgb}{0.9, 0.89, 0.89}
\definecolor{plum(traditional)}{rgb}{0.56, 0.27, 0.52}
\definecolor{plum(web)}{rgb}{0.8, 0.6, 0.8}
\definecolor{portlandorange}{rgb}{1.0, 0.35, 0.21}
\definecolor{powderblue(web)}{rgb}{0.69, 0.88, 0.9}
\definecolor{princetonorange}{rgb}{1.0, 0.56, 0.0}
\definecolor{prune}{rgb}{0.44, 0.11, 0.11}
\definecolor{prussianblue}{rgb}{0.0, 0.19, 0.33}
\definecolor{psychedelicpurple}{rgb}{0.87, 0.0, 1.0}
\definecolor{puce}{rgb}{0.8, 0.53, 0.6}
\definecolor{pumpkin}{rgb}{1.0, 0.46, 0.09}
\definecolor{purple(html/css)}{rgb}{0.5, 0.0, 0.5}
\definecolor{purple(munsell)}{rgb}{0.62, 0.0, 0.77}
\definecolor{purple(x11)}{rgb}{0.63, 0.36, 0.94}
\definecolor{purpleheart}{rgb}{0.41, 0.21, 0.61}
\definecolor{purplemountainmajesty}{rgb}{0.59, 0.47, 0.71}
\definecolor{purplepizzazz}{rgb}{1.0, 0.31, 0.85}
\definecolor{purpletaupe}{rgb}{0.31, 0.25, 0.3}
\definecolor{radicalred}{rgb}{1.0, 0.21, 0.37}
\definecolor{raspberry}{rgb}{0.89, 0.04, 0.36}
\definecolor{raspberryglace}{rgb}{0.57, 0.37, 0.43}
\definecolor{raspberrypink}{rgb}{0.89, 0.31, 0.61}
\definecolor{raspberryrose}{rgb}{0.7, 0.27, 0.42}
\definecolor{rawumber}{rgb}{0.51, 0.4, 0.27}
\definecolor{razzledazzlerose}{rgb}{1.0, 0.2, 0.8}
\definecolor{razzmatazz}{rgb}{0.89, 0.15, 0.42}
\definecolor{red}{rgb}{1.0, 0.0, 0.0}
\definecolor{red(munsell)}{rgb}{0.95, 0.0, 0.24}
\definecolor{red(ncs)}{rgb}{0.77, 0.01, 0.2}
\definecolor{red(pigment)}{rgb}{0.93, 0.11, 0.14}
\definecolor{red(ryb)}{rgb}{1.0, 0.15, 0.07}
\definecolor{red-brown}{rgb}{0.65, 0.16, 0.16}
\definecolor{red-violet}{rgb}{0.78, 0.08, 0.52}
\definecolor{redwood}{rgb}{0.67, 0.31, 0.32}
\definecolor{regalia}{rgb}{0.32, 0.18, 0.5}
\definecolor{richblack}{rgb}{0.0, 0.25, 0.25}
\definecolor{richbrilliantlavender}{rgb}{0.95, 0.65, 1.0}
\definecolor{richcarmine}{rgb}{0.84, 0.0, 0.25}
\definecolor{richelectricblue}{rgb}{0.03, 0.57, 0.82}
\definecolor{richlavender}{rgb}{0.67, 0.38, 0.8}
\definecolor{richlilac}{rgb}{0.71, 0.4, 0.82}
\definecolor{richmaroon}{rgb}{0.69, 0.19, 0.38}
\definecolor{riflegreen}{rgb}{0.25, 0.28, 0.2}
\definecolor{robineggblue}{rgb}{0.0, 0.8, 0.8}
\definecolor{rose}{rgb}{1.0, 0.0, 0.5}
\definecolor{rosebonbon}{rgb}{0.98, 0.26, 0.62}
\definecolor{roseebony}{rgb}{0.4, 0.3, 0.28}
\definecolor{rosegold}{rgb}{0.72, 0.43, 0.47}
\definecolor{rosemadder}{rgb}{0.89, 0.15, 0.21}
\definecolor{rosepink}{rgb}{1.0, 0.4, 0.8}
\definecolor{rosequartz}{rgb}{0.67, 0.6, 0.66}
\definecolor{rosetaupe}{rgb}{0.56, 0.36, 0.36}
\definecolor{rosevale}{rgb}{0.67, 0.31, 0.32}
\definecolor{rosewood}{rgb}{0.4, 0.0, 0.04}
\definecolor{rossocorsa}{rgb}{0.83, 0.0, 0.0}
\definecolor{rosybrown}{rgb}{0.74, 0.56, 0.56}
\definecolor{royalazure}{rgb}{0.0, 0.22, 0.66}
\definecolor{royalblue(traditional)}{rgb}{0.0, 0.14, 0.4}
\definecolor{royalblue(web)}{rgb}{0.25, 0.41, 0.88}
\definecolor{royalfuchsia}{rgb}{0.79, 0.17, 0.57}
\definecolor{royalpurple}{rgb}{0.47, 0.32, 0.66}
\definecolor{ruby}{rgb}{0.88, 0.07, 0.37}
\definecolor{ruddy}{rgb}{1.0, 0.0, 0.16}
\definecolor{ruddybrown}{rgb}{0.73, 0.4, 0.16}
\definecolor{ruddypink}{rgb}{0.88, 0.56, 0.59}
\definecolor{rufous}{rgb}{0.66, 0.11, 0.03}
\definecolor{russet}{rgb}{0.5, 0.27, 0.11}
\definecolor{rust}{rgb}{0.72, 0.25, 0.05}
\definecolor{sacramentostategreen}{rgb}{0.0, 0.34, 0.25}
\definecolor{saddlebrown}{rgb}{0.55, 0.27, 0.07}
\definecolor{safetyorange(blazeorange)}{rgb}{1.0, 0.4, 0.0}
\definecolor{saffron}{rgb}{0.96, 0.77, 0.19}
\definecolor{st.patricksblue}{rgb}{0.14, 0.16, 0.48}
\definecolor{salmon}{rgb}{1.0, 0.55, 0.41}
\definecolor{salmonpink}{rgb}{1.0, 0.57, 0.64}
\definecolor{sand}{rgb}{0.76, 0.7, 0.5}
\definecolor{sanddune}{rgb}{0.59, 0.44, 0.09}
\definecolor{sandstorm}{rgb}{0.93, 0.84, 0.25}
\definecolor{sandybrown}{rgb}{0.96, 0.64, 0.38}
\definecolor{sandytaupe}{rgb}{0.59, 0.44, 0.09}
\definecolor{sangria}{rgb}{0.57, 0.0, 0.04}
\definecolor{sapgreen}{rgb}{0.31, 0.49, 0.16}
\definecolor{sapphire}{rgb}{0.03, 0.15, 0.4}
\definecolor{satinsheengold}{rgb}{0.8, 0.63, 0.21}
\definecolor{scarlet}{rgb}{1.0, 0.13, 0.0}
\definecolor{schoolbusyellow}{rgb}{1.0, 0.85, 0.0}
\definecolor{screamingreen}{rgb}{0.46, 1.0, 0.44}
\definecolor{seagreen}{rgb}{0.18, 0.55, 0.34}
\definecolor{sealbrown}{rgb}{0.2, 0.08, 0.08}
\definecolor{seashell}{rgb}{1.0, 0.96, 0.93}
\definecolor{selectiveyellow}{rgb}{1.0, 0.73, 0.0}
\definecolor{sepia}{rgb}{0.44, 0.26, 0.08}
\definecolor{shadow}{rgb}{0.54, 0.47, 0.36}
\definecolor{shamrockgreen}{rgb}{0.0, 0.62, 0.38}
\definecolor{shockingpink}{rgb}{0.99, 0.06, 0.75}
\definecolor{sienna}{rgb}{0.53, 0.18, 0.09}
\definecolor{silver}{rgb}{0.75, 0.75, 0.75}
\definecolor{sinopia}{rgb}{0.8, 0.25, 0.04}
\definecolor{skobeloff}{rgb}{0.0, 0.48, 0.45}
\definecolor{skyblue}{rgb}{0.53, 0.81, 0.92}
\definecolor{skymagenta}{rgb}{0.81, 0.44, 0.69}
\definecolor{slateblue}{rgb}{0.42, 0.35, 0.8}
\definecolor{slategray}{rgb}{0.44, 0.5, 0.56}
\definecolor{smalt(darkpowderblue)}{rgb}{0.0, 0.2, 0.6}
\definecolor{smokeytopaz}{rgb}{0.58, 0.25, 0.03}
\definecolor{smokyblack}{rgb}{0.06, 0.05, 0.03}
\definecolor{snow}{rgb}{1.0, 0.98, 0.98}
\definecolor{spirodiscoball}{rgb}{0.06, 0.75, 0.99}
\definecolor{splashedwhite}{rgb}{1.0, 0.99, 1.0}
\definecolor{springbud}{rgb}{0.65, 0.99, 0.0}
\definecolor{springgreen}{rgb}{0.0, 1.0, 0.5}
\definecolor{steelblue}{rgb}{0.27, 0.51, 0.71}
\definecolor{stildegrainyellow}{rgb}{0.98, 0.85, 0.37}
\definecolor{straw}{rgb}{0.89, 0.85, 0.44}
\definecolor{sunglow}{rgb}{1.0, 0.8, 0.2}
\definecolor{sunset}{rgb}{0.98, 0.84, 0.65}
\definecolor{tan}{rgb}{0.82, 0.71, 0.55}
\definecolor{tangelo}{rgb}{0.98, 0.3, 0.0}
\definecolor{tangerine}{rgb}{0.95, 0.52, 0.0}
\definecolor{tangerineyellow}{rgb}{1.0, 0.8, 0.0}
\definecolor{taupe}{rgb}{0.28, 0.24, 0.2}
\definecolor{taupegray}{rgb}{0.55, 0.52, 0.54}
\definecolor{teagreen}{rgb}{0.82, 0.94, 0.75}
\definecolor{tearose(orange)}{rgb}{0.97, 0.51, 0.47}
\definecolor{tearose(rose)}{rgb}{0.96, 0.76, 0.76}
\definecolor{teal}{rgb}{0.0, 0.5, 0.5}
\definecolor{tealblue}{rgb}{0.21, 0.46, 0.53}
\definecolor{tealgreen}{rgb}{0.0, 0.51, 0.5}
\definecolor{tawny}{rgb}{0.8, 0.34, 0.0}
\definecolor{terracotta}{rgb}{0.89, 0.45, 0.36}
\definecolor{thistle}{rgb}{0.85, 0.75, 0.85}
\definecolor{thulianpink}{rgb}{0.87, 0.44, 0.63}
\definecolor{ticklemepink}{rgb}{0.99, 0.54, 0.67}
\definecolor{tiffanyblue}{rgb}{0.04, 0.73, 0.71}
\definecolor{tigerseye}{rgb}{0.88, 0.55, 0.24}
\definecolor{timberwolf}{rgb}{0.86, 0.84, 0.82}
\definecolor{titaniumyellow}{rgb}{0.93, 0.9, 0.0}
\definecolor{tomato}{rgb}{1.0, 0.39, 0.28}
\definecolor{toolbox}{rgb}{0.45, 0.42, 0.75}
\definecolor{tractorred}{rgb}{0.99, 0.05, 0.21}
\definecolor{trolleygrey}{rgb}{0.5, 0.5, 0.5}
\definecolor{tropicalrainforest}{rgb}{0.0, 0.46, 0.37}
\definecolor{trueblue}{rgb}{0.0, 0.45, 0.81}
\definecolor{tuftsblue}{rgb}{0.28, 0.57, 0.81}
\definecolor{tumbleweed}{rgb}{0.87, 0.67, 0.53}
\definecolor{turkishrose}{rgb}{0.71, 0.45, 0.51}
\definecolor{turquoise}{rgb}{0.19, 0.84, 0.78}
\definecolor{turquoiseblue}{rgb}{0.0, 1.0, 0.94}
\definecolor{turquoisegreen}{rgb}{0.63, 0.84, 0.71}
\definecolor{tuscanred}{rgb}{0.51, 0.21, 0.21}
\definecolor{twilightlavender}{rgb}{0.54, 0.29, 0.42}
\definecolor{tyrianpurple}{rgb}{0.4, 0.01, 0.24}
\definecolor{uablue}{rgb}{0.0, 0.2, 0.67}
\definecolor{uared}{rgb}{0.85, 0.0, 0.3}
\definecolor{ube}{rgb}{0.53, 0.47, 0.76}
\definecolor{uclablue}{rgb}{0.33, 0.41, 0.58}
\definecolor{uclagold}{rgb}{1.0, 0.7, 0.0}
\definecolor{ufogreen}{rgb}{0.24, 0.82, 0.44}
\definecolor{ultramarine}{rgb}{0.07, 0.04, 0.56}
\definecolor{ultramarineblue}{rgb}{0.25, 0.4, 0.96}
\definecolor{ultrapink}{rgb}{1.0, 0.44, 1.0}
\definecolor{umber}{rgb}{0.39, 0.32, 0.28}
\definecolor{unitednationsblue}{rgb}{0.36, 0.57, 0.9}
\definecolor{unmellowyellow}{rgb}{1.0, 1.0, 0.4}
\definecolor{upforestgreen}{rgb}{0.0, 0.27, 0.13}
\definecolor{upmaroon}{rgb}{0.48, 0.07, 0.07}
\definecolor{upsdellred}{rgb}{0.68, 0.09, 0.13}
\definecolor{urobilin}{rgb}{0.88, 0.68, 0.13}
\definecolor{usccardinal}{rgb}{0.6, 0.0, 0.0}
\definecolor{uscgold}{rgb}{1.0, 0.8, 0.0}
\definecolor{utahcrimson}{rgb}{0.83, 0.0, 0.25}
\definecolor{vanilla}{rgb}{0.95, 0.9, 0.67}
\definecolor{vegasgold}{rgb}{0.77, 0.7, 0.35}
\definecolor{venetianred}{rgb}{0.78, 0.03, 0.08}
\definecolor{verdigris}{rgb}{0.26, 0.7, 0.68}
\definecolor{vermilion}{rgb}{0.89, 0.26, 0.2}
\definecolor{veronica}{rgb}{0.63, 0.36, 0.94}
\definecolor{violet}{rgb}{0.56, 0.0, 1.0}
\definecolor{violet(colorwheel)}{rgb}{0.5, 0.0, 1.0}
\definecolor{violet(ryb)}{rgb}{0.53, 0.0, 0.69}
\definecolor{violet(web)}{rgb}{0.93, 0.51, 0.93}
\definecolor{viridian}{rgb}{0.25, 0.51, 0.43}
\definecolor{vividauburn}{rgb}{0.58, 0.15, 0.14}
\definecolor{vividburgundy}{rgb}{0.62, 0.11, 0.21}
\definecolor{vividcerise}{rgb}{0.85, 0.11, 0.51}
\definecolor{vividtangerine}{rgb}{1.0, 0.63, 0.54}
\definecolor{vividviolet}{rgb}{0.62, 0.0, 1.0}
\definecolor{warmblack}{rgb}{0.0, 0.26, 0.26}
\definecolor{wenge}{rgb}{0.39, 0.33, 0.32}
\definecolor{wheat}{rgb}{0.96, 0.87, 0.7}
\definecolor{white}{rgb}{1.0, 1.0, 1.0}
\definecolor{whitesmoke}{rgb}{0.96, 0.96, 0.96}
\definecolor{wildblueyonder}{rgb}{0.64, 0.68, 0.82}
\definecolor{wildstrawberry}{rgb}{1.0, 0.26, 0.64}
\definecolor{wildwatermelon}{rgb}{0.99, 0.42, 0.52}
\definecolor{wisteria}{rgb}{0.79, 0.63, 0.86}
\definecolor{xanadu}{rgb}{0.45, 0.53, 0.47}
\definecolor{yaleblue}{rgb}{0.06, 0.3, 0.57}
\definecolor{yellow}{rgb}{1.0, 1.0, 0.0}
\definecolor{yellow(munsell)}{rgb}{0.94, 0.8, 0.0}
\definecolor{yellow(ncs)}{rgb}{1.0, 0.83, 0.0}
\definecolor{yellow(process)}{rgb}{1.0, 0.94, 0.0}
\definecolor{yellow(ryb)}{rgb}{1.0, 1.0, 0.2}
\definecolor{yellow-green}{rgb}{0.6, 0.8, 0.2}
\definecolor{zaffre}{rgb}{0.0, 0.08, 0.66}
\definecolor{zinnwalditebrown}{rgb}{0.17, 0.09, 0.03}

\newcommand{\eff}{\text{eff}}

\newcommand{\emma}[1]{#1}

\begin{document}

\title{The contact theorem for charged fluids: from planar to curved geometries} 

\author{Juan Pablo Mallarino}
\author{Gabriel  T\'ellez}
\affiliation{Departamento de F\'{\i}sica, Universidad de los Andes, Bogotá, Colombia}
\author{ Emmanuel Trizac}
\affiliation{Universit\'e Paris-Sud, Laboratoire de Physique Th\'eorique et 
Mod\`eles Statistiques, UMR CNRS 8626, 91405 Orsay, France}

\begin{abstract}
When a Coulombic fluid is confined between two parallel charged plates,
an exact relation links the difference of ionic densities at contact
with the plates, to the surface charges of these boundaries.
It no longer applies when the boundaries are curved, and we
work out how it generalizes when the fluid is confined between 
two concentric spheres (or cylinders), in two and in three space dimensions.
The analysis is thus performed within the cell model picture. The generalized contact relation opens 
the possibility to derive new exact expressions, of particular interest in the regime
of strong coulombic couplings. Some emphasis is put on cylindrical geometry,
for which we discuss in depth the phenomenon of counter-ion evaporation/condensation,
and obtain novel results.
Good agreement is found with Monte Carlo simulation data.
\end{abstract}

\date{\today}
\maketitle

\section{Introduction}

Exact results in the equilibrium statistical mechanics of charged fluids are scarce \cite{HL00,Levin02,Belloni00,Messina09},
leaving aside the formal body of relations that connect quantities that 
cannot be obtained explicitly. Most often, the exact results pertain to two
dimensional systems \cite{Janco81,Forrester98,Samaj03}, where charges 
interact by a logarithmic potential.
Yet, an interesting and useful exact relation is provided by the so-called contact theorem \cite{contact1,contact2,contact3},
that is not limited to two dimensions, in the sense that it also applies
when charges interact through a $1/r$ potential as is the case
in three dimensions. To provide an insight, we introduce
the Bjerrum length $\ell_B$, to be defined below from the temperature and the solvent
permittivity (treated as a dielectric continuum); the contact theorem
holds for charges, point-like or with a given hard-core, 
confined between two parallel planar structureless interfaces, having
respective surface charges $\sigma_a e$ and $\sigma_b e$, where $e$ is the elementary charge. It simply relates 
the pressure $P$ to the contact densities of ions ($n_a$ for the total
ionic concentration in contact with plate $a$):
\begin{equation}
\beta P \, =\, n_a - 2 \pi \ell_B \sigma_a^2,
\end{equation}
where $\beta = 1/(kT)$ is proportional to the inverse temperature.
A similar relation holds at contact with plate $b$ where the total density is
$n_b$:
\begin{equation}
\beta P \, =\, n_b - 2 \pi \ell_B \sigma_b^2.
\end{equation}
This implies that for uncharged walls, we have $\beta P = n_a=n_b$,
which provides an exact (although not explicit) equation of state for a hard sphere
fluid (see e.g. \cite{AJP08}).
Another limiting case of more significance to us is obtained when the distance between the two plates (also referred
to later as the macro-ions) diverges,
which leads to a vanishing pressure, and thus to an exact constraint
between contact density and surface charge. This allows to discriminate
various approximate approaches \cite{FrLe12,rque10}. In other circumstances, 
knowing the ionic density profile between the charged plates, one can 
infer the equation of state. This is the route followed in the strong
coupling analysis of Refs \cite{Netz01,WSC1,WSC2}, where an exact 
and explicit equation of state can be obtained at short distances
\cite{Varenna}.

However, the exact planar relation \cite{rque5}
\begin{equation}
 n_a - 2 \pi \ell_B \,\sigma_a^2 \, = \, n_b - 2 \pi \ell_B \,\sigma_b^2
\label{eq:contact_planar}
\end{equation}
breaks down as soon as the charged interfaces are no longer planar but bear some curvature. This is
regrettable since knowing the counterpart of Eq. \eqref{eq:contact_planar}
would be desirable for analytical progress, as well as for testing numerical
simulations. Our main motivation is to fill this gap. To this end, we shall work 
in the framework of the cell model \cite{Fuoss01091951,Marcus55,DeHo01,Levin02}, where a charged body (cylindrical or spherical,
and bearing in the following the subscript $a$),
is enclosed in a concentric (Wigner-Seitz) cell of a similar shape (referred to with subscript $b$), and we shall analyze the fate 
of the incorrect planar relation \eqref{eq:contact_planar}.
The cell model approach has proven fruitful and provides accurate results 
for quantities such as the pressure, that can be compared against experiments and numerical simulations
\cite{LoLi99,BTA_JCP_02,levin_tb03,TeTr03,Antypov06,Claudio09,Denton,Jonsson12}. Its interest is that it is in essence a one macro-ion
approach, and thus considerably simpler than the original full $N$-macro-ion problem.

At this point, a clarification is in order. Within the cell model, and thus with curved
macro-ions, an exact result holds \cite{contact3,DeHo01},  
\begin{equation}
\beta P \, =\, n_b - 2 \pi \ell_B \sigma_b^2,
\label{eq:pressure_contact}
\end{equation}
where by definition $b$ denotes the outer boundary (see Fig. \ref{fig:art_sph} below).
This relation is often particularized to the case $\sigma_b=0$ \cite{rque19}, and relevant
in numerical simulations to get the pressure from the ionic density at contact 
with the confining boundary \cite{rque20}. In all our analysis, Eq. 
\eqref{eq:pressure_contact} will remain valid, but will not be of particular
interest (apart from allowing to introduce the pressure in relations
where it does not explicitly appear). Our interest instead goes to finding
the connection between $n_a$, $n_b$, $\sigma_a$ and $\sigma_b$, which should
reduce to \eqref{eq:contact_planar} when curvatures vanish.

The outline of the paper is as follows. The model is laid out in section 
\ref{sec:def}, where a generalized contact theorem is derived. 
Different geometries must be distinguished, and will shall consider
explicitly three different cases: a sphere within a confining sphere
(in two dimensions with a log potential, or in three dimension
with a $1/r$ potential, see Fig. \ref{fig:art_sph}), together with a cylinder within a cylinder.
In two dimensions, the latter case is equivalent to the previous
2d spherical problem (a charged disc within a disc, see Fig. \ref{fig:cyl_artistic}), so that
only the 3D case is of interest here. 
In section \ref{sec:explicit_then_planar}, the previous formal relations 
will be made more explicit, whenever possible, and it will be shown that 
upon taking the planar limit in a suitable fashion, one recovers the known 
relation \eqref{eq:contact_planar}. The remainder of the paper will
be devoted to discussing practical consequences of the generalized contact
relation: first considering cylindrical macro-ions (in two or three dimensions)
in section \ref{sec:applI}, and then spherical macro-ions in section 
\ref{sec:applII}. In section \ref{sec:applI}, our analytical results will be compared
to measures performed in Monte Carlo simulations, following the centrifugal scheme
used in \cite{NN06_PRE,MTT13}, to which the reader is referred for further details.
Particular attention will be paid to the weakly as well and strongly coupled regimes.

\begin{figure}[h!]
\centering
\subfloat[][2d]{\includegraphics[width=.3\linewidth,angle=90]{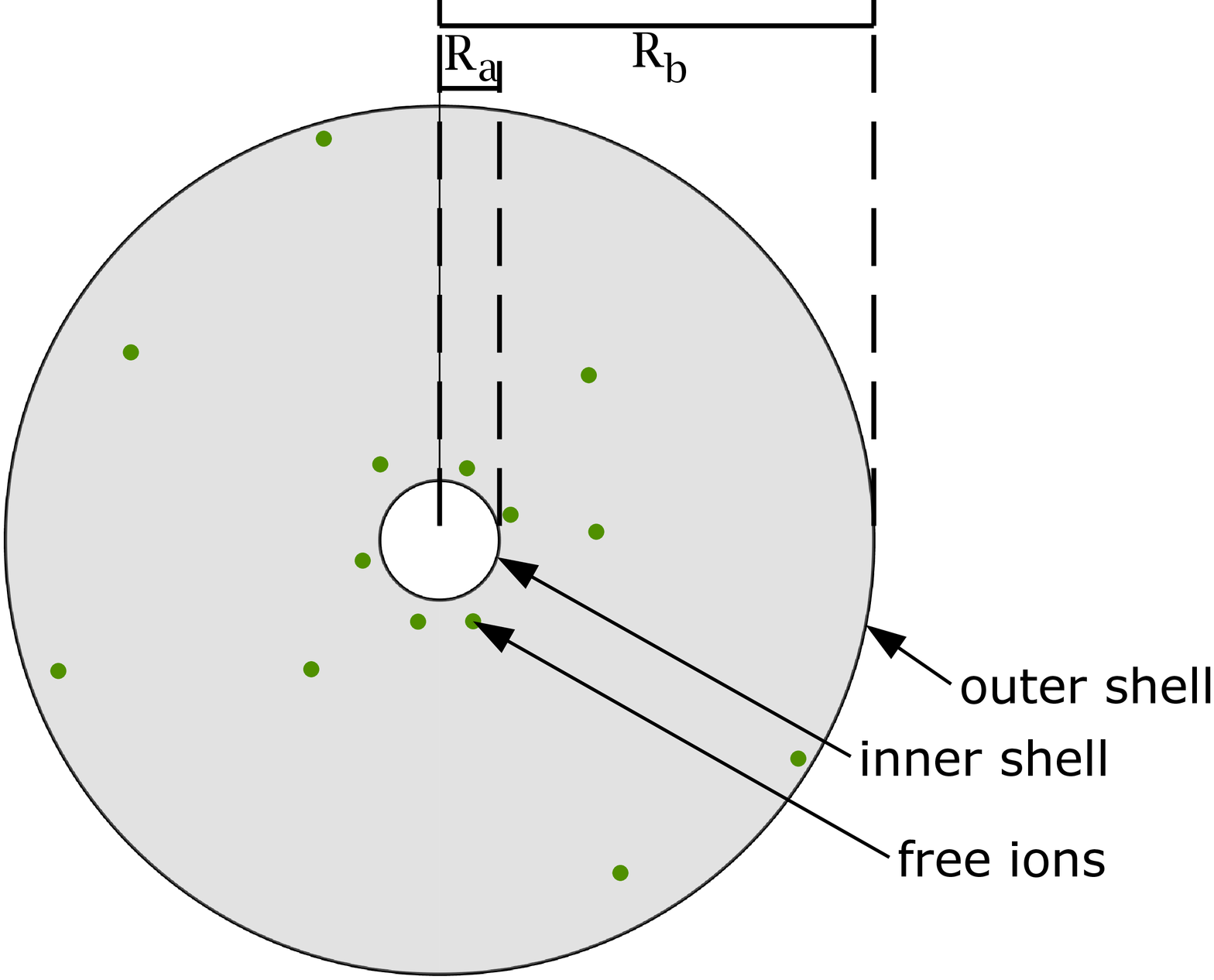}}%
\quad%
\subfloat[][3d]{\includegraphics[width=.3\linewidth,angle=0]{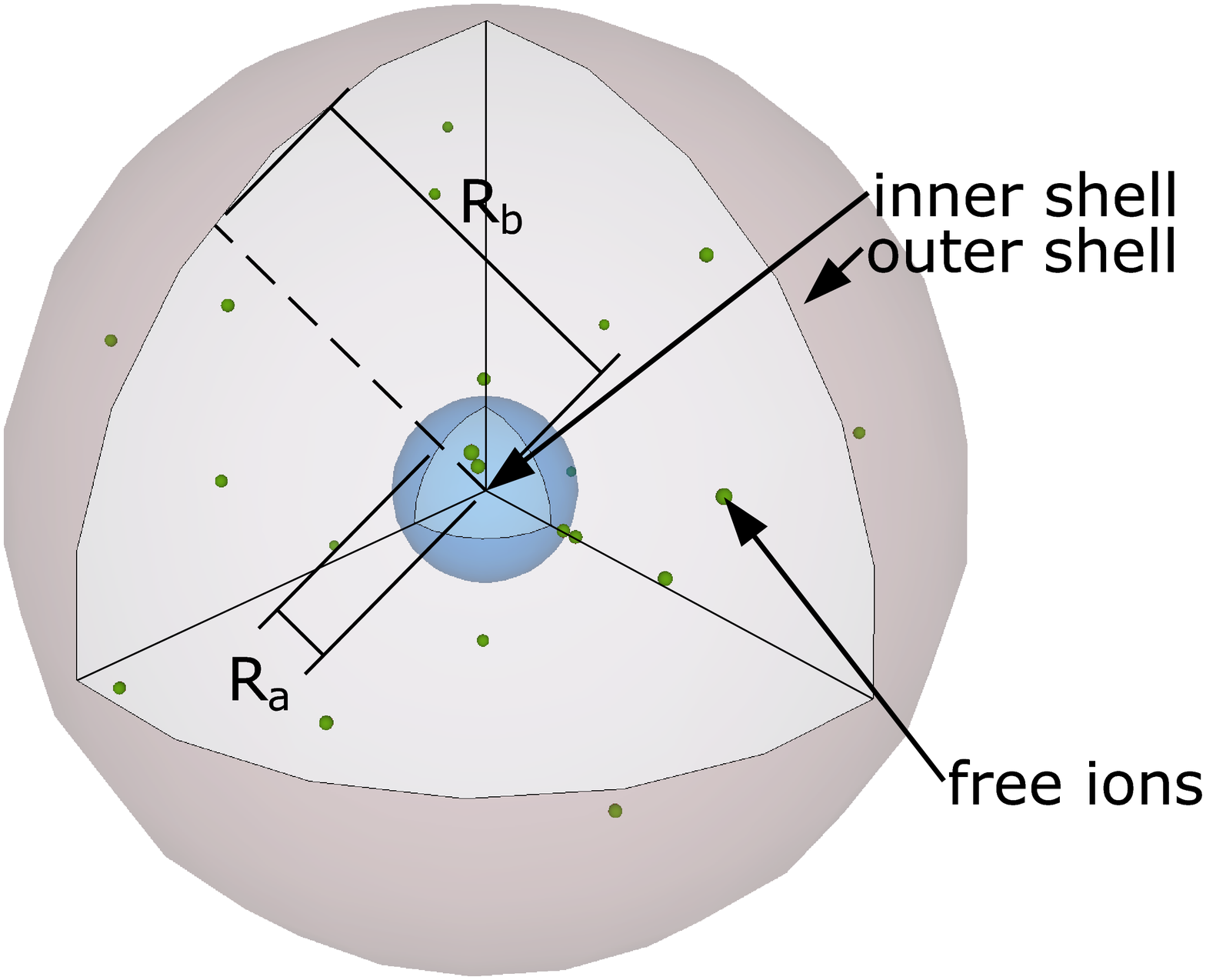}}%
\quad%
\caption[]{Schematic view of the spherical cell in two and three dimensions. The charged macro-ion has radius 
$R_a$ while $R_b$ denotes the size of the confining cell.}
\label{fig:art_sph}
\end{figure}

\ImTR{0.3}{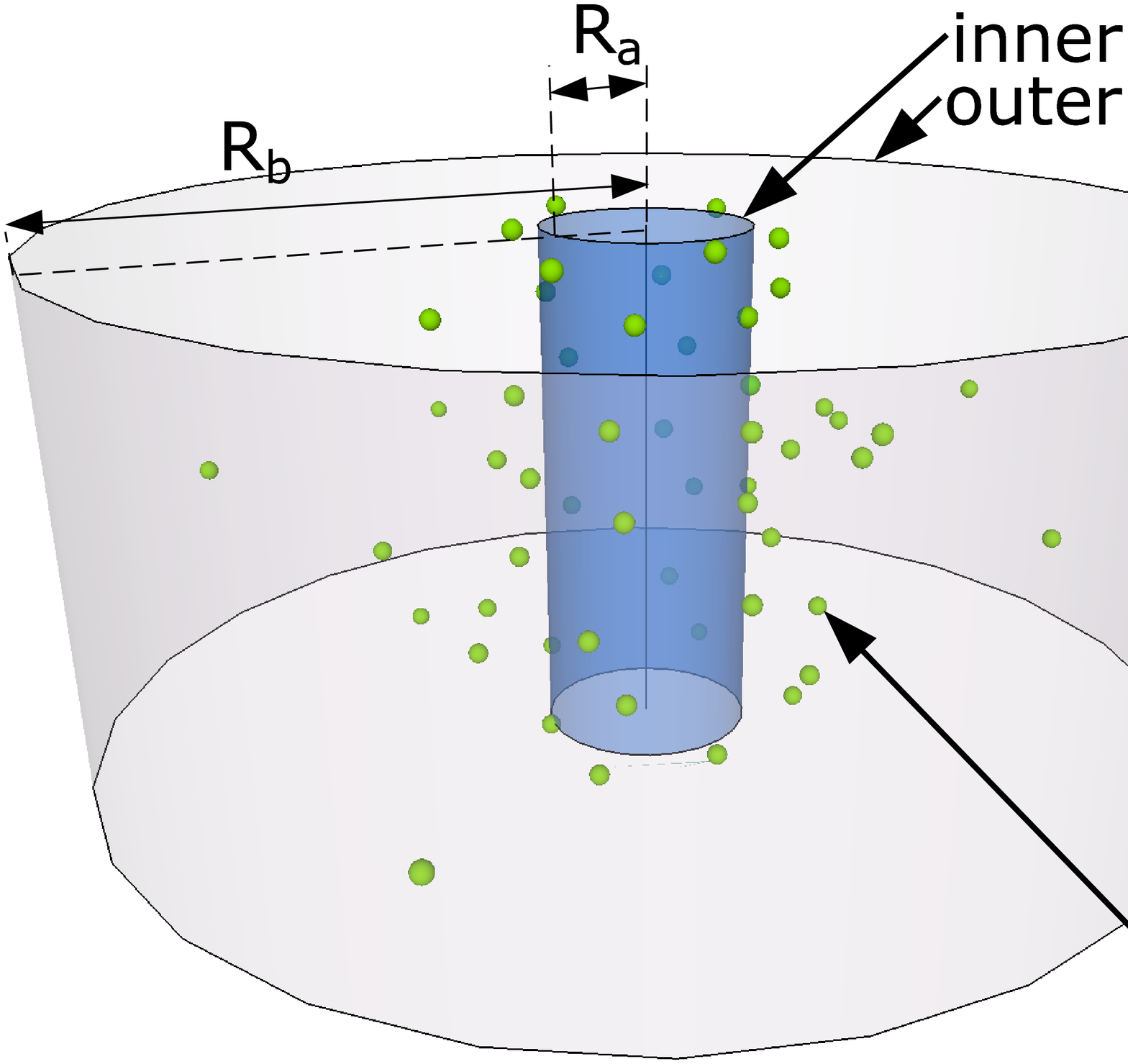}{3D_cyl}{Schematic view of the cylindrical cell in three
dimensions. The system is infinite along the cylinders' axis. Such a setup is commonplace and relevant
to the study of charged polyelectrolytes \cite{Fuoss01091951,DeHo01}.
The two dimensional case is identical to Fig. \ref{fig:art_sph} (a).
\label{fig:cyl_artistic}}{0}

\section{Definitions and derivation of a formal contact relation}
\label{sec:def}

\subsection{The Hamiltonian and the virial}

We consider $N$ charged particles (with charges $q_{i}e$), $e$ being the elementary charge and $q_i$ the valency, that occupy the domain $\Lambda$ between two concentric
shells, with radius $R_{a}<R_{b}$. The frontier of the domain is denoted
$\partial\Lambda$. Each shell may carry a total charge
$Q_{a}e$ and $Q_{b}e$, with uniform surface charge densities $\sigma_a e$
and $\sigma_b e$. The system is globally neutral $\sum_{i} q_{i}
+Q_{a}+Q_{b}=0$, at equilibrium with temperature $1/\beta$.
The Hamiltonian of the system reads
\begin{equation}
  \label{eq:h}
  H= e^2\sum_{i<j} q_i q_j v(r_{ij}) + e^2\sum_{i} q_{i} v_a(r_i) 
  + e^2\sum_{i} q_{i} v_b + e^2 V_a + e^2 V_b +e^2 V_{ab}
\end{equation}
where $v(r)$ is the Coulomb pair potential, $v_a(r)$ the potential
created by the inner shell, $v_b$ the potential created by the outer
shell (constant), $V_a$ and $V_b$ the self-energy of each shell, and
$V_{ab}$ the interaction potential between the shells.
Making use of the global electro-neutrality of the system and of the explicit expressions
recapitulated in Table \ref{table:1}, we get 
\begin{equation}
  \label{eq:h3d}
  H / e^2=\sum_{i} q_{i} v_a(r_i)+\sum_{i<j} q_i q_j v(r_{ij})
  +V(R_a,R_b)
  \,,
\end{equation}
with, for a 3D system,
\begin{equation}
V(R_a,R_b) \,=\, \frac{Q_{a}^2}{2R_{a}}  - \frac{Q_{b}^2}{2R_{b}} 
\end{equation}
while the corresponding 2d results follow from the replacement 
$1/r \to -\log r$:
\begin{equation}
  V(R_a,R_b) \,=\, -\frac{Q_{a}^2}{2} \ln R_{a}  + \frac{Q_{b}^2}{2} \ln
  R_{b} 
  \,.
\end{equation}
We did not include a possible hard core exclusion between the ions, and between 
the ions and the shells (interfaces), for it is rather immaterial for the subsequent
discussion, and does not affect the results. Thus, whenever a 'contact' density will be referred to, it should be understood
that it pertains to the distance of closest approach between two charged bodies,
and not to physically vanishing distances.

\begin{table}[htb]
  \centering
  \begin{tabular}{|c|c|c|c|c|c|c|}
    \hline
     \null       & $v(r)$ & $v_a(r)$ & $v_b$ & $V_a$& $V_b$  & $V_{ab}$\\ \hline
    $d=2$   & $-\ln r$      & $-Q_a \ln r$    & $-Q_b \ln R_b$   & $-{Q_a^2}(\ln R_a)/2$& $-{Q_b^2}(\ln R_b)/2$ & $-Q_{a} Q_{b} \ln R_{b}$ \\
    $d=3$   & $1/r$ & $ {Q_a}/{r}$ &${Q_b}/{R_b}$ & $ {Q_a^2}/{(2R_a)}$ & ${Q_b^2}/{(2R_b)}$ & ${Q_{a} Q_{b}}/{R_{b}}$ \\
    \hline
  \end{tabular}
  \caption{Dependence of the various one-body or two-body potentials on space dimension $d$.}
  \label{table:1}
\end{table}

To avoid cumbersome expressions, the dielectric permittivity is not included in the 
Hamiltonian, and is set to unity. To prevent possible confusions, we shall make use of the often employed
coupling constant $\Gamma = \beta e^2$ for two dimensional systems, and in three
dimensions, of the 
Bjerrum length $\ell_B = \beta e^2/\varepsilon$, where $\varepsilon$ is the permittivity of the medium (solvent).
In water at room temperature, $\ell_B \simeq 0.7\,$nm.

In both cases, the configurational partition function can be
written as
\begin{equation}
  Z=Z^{*} e^{-\beta V(R_a,R_b)}
\end{equation}
with
\begin{equation}
  \label{eq:Zstar}
  Z^{*} = \int_{\Lambda^N} d\mathbf{r}^{N} \exp\left[-\beta e^2 \left( \sum_{i<j} 
    q_i q_j v(\mathbf{r}_{ij}) 
    +\sum_{i} q_{i} v_a (\mathbf{r}_{i}) \right) \right].
\end{equation}
In the following discussion, the virial $W$ will be an important
quantity. It is the sum of one-body and two-body terms and is defined as
\begin{equation}
  \label{eq:W}
  W / e^2=- \sum_{i<j} q_{i}q_{j} \mathbf{r}_{ij} \cdot\frac{d v}{d \mathbf{r}_{ij}} (\mathbf{r}_{ij})
  -\sum_{i} q_i \mathbf{r}_{i} \cdot\frac{d v_{a}}{d\mathbf{r}_i}(r_{i})
  \,.
\end{equation}
Since 
$
  \label{eq:vir-r}
  r \frac{d}{dr}\left(\frac{1}{r}\right)=-\frac{1}{r}
  \,,
  $
we have in three dimensions that
\begin{equation}
  \label{eq:W-H}
  W/e^2 = H/e^2 -  \frac{Q_{a}^2}{2R_{a}}  + \frac{Q_{b}^2}{2R_{b}} 
\end{equation}
while, in two dimensions:
\begin{equation}
  \label{eq:w2d-gen}
  W /e^2=  \frac{1}{2} \sum_{i}\sum_{j\neq i} q_i q_j + Q_a \sum_{i} q_i
\end{equation}
Taking into account electro-neutrality, the latter expression yields
\begin{equation}
  \label{eq:w2d}
  W /e^2 =  \frac{1}{2} \left[ Q_{b}^2- Q_{a}^2
    -\sum_{i} q_{i}^2\right] .
\end{equation}
In particular, if there is only one species of charged particles,
$q_{i}=q$, electro-neutrality reads $qN+Q_a+Q_b=0$, and $W$ takes a 
particularly simple form
\begin{equation}
  \label{eq:W2d2}
  W/e^2 =  \frac{1}{2} \left[ Q_{b}^2- Q_{a}^2
    +q (Q_a+Q_b) \right].
\end{equation}

\subsection{Derivation of the generalized contact theorem}
\label{sec:sum-rule}

\subsubsection{Spherical geometry ($d=2$ and $d=3$)}

We aim at getting the pressure of the system, though the volume derivative 
of its free energy.
Let us compute the derivative of the partition function with respect
to $R_b$, with fixed $R_a$, $Q_a$, $Q_b$, and $N$. Let ${\cal V}_b=4\pi
R_{b}^{3}/3$ (3D) or ${\cal V}_b=\pi R_{b}^2$ (2D) be the volume (area)
enclosed by the cell.
We have
\begin{equation}
  \label{eq:pVab}
  \frac{dV(R_a,R_b)}{d{\cal V}_b}
  =\frac{\sigma_b^2}{2\chi}
\end{equation}
with $\chi=1/(4\pi)$ in three dimensions, $\chi=1/(2\pi)$
in two dimensions, and $\sigma_{b}$ is the surface charge density at
the outer shell. In 3D, $Q_i = 4\pi R_i^2 \sigma_i$ while in 2D,
the $\sigma_i$ with $i=a$ or $i=b$ have the meaning of a line charge: $Q_i = 2\pi R_i \sigma_i$.

An explicit derivation of $Z^{*}$ with respect to $R_b$ reduces the
$N$-multiple integrals to $(N-1)$-multiple integrals with the position
of one particle fixed at $r=R_b$, see e.g. \cite{DeHo01}, thus giving a term directly related
to the density at $r=R_b$:
\begin{equation}
  \label{eq:p1}
  \frac{\partial \ln Z^{*}}{\partial {\cal V}_b} = 
  n(R_{b})
\end{equation}
where $n(R_{b})$ is the total density at the edge of the cell. In the
case of a multicomponent system, it is the sum of the densities of each
species $n(R_{b})=\sum_{\alpha} n_{\alpha}(R_{b})$.

Alternatively, the derivative can be computed using the following
scaling argument. In the configurational integral, we make the change
of variable $r=\widetilde{r} R_{b}$, such that the upper limit of
integration is 1. On the other hand, the lower limit of integration
depends on $R_{b}$, since it is now $R_{a}/R_{b}$. Also, the Boltzmann
factor in the integral now depends on $R_b$:
\begin{equation}
  Z^{*}=R^{N d}
  \int \prod_{i} d\Omega_{i} 
  \int_{[R_a/R_b,1]^{N}} \prod_{i} \widetilde{r}_i^{d-1}d\widetilde{r}_{i}
  \exp\left[-\beta e^2\left( \sum_{i<j} 
    q_i q_j v(R_{b} \widetilde{r}_{ij}) 
    +\sum_{i} q_{i} v_a (R_{b} \widetilde {r}_{i}) \right) \right]
\end{equation} 
where $d=2,3$ is the dimension and $\Omega_i$ corresponds to the solid angle. Taking the derivative with respect to
${\cal V}_b$, gives
\begin{equation}
  \label{eq:dZ}
  \frac{\partial \ln Z^{*}}{\partial {\cal V}_b}
  = \frac{N}{{\cal V}_b} + \frac{1}{{\cal V}_b \, d} \langle \beta W \rangle
  +
  \frac{{\cal V}_a}{{\cal V}_b} n(R_a)
  \,,
\end{equation}
where the brackets $\langle ...\rangle$ denote statistical (canonical) average.
Therefore, we have the relation,
\begin{equation}
  \label{eq:cont1}
  {\cal V}_{b} n(R_b)=
  N +  \frac{1}{d} \langle \beta W \rangle
  +{\cal V}_{a} n(R_a)
  \,.
\end{equation}
We note in passing that this relation can also be obtained from
application of the virial theorem $\langle 2T + {\cal W}\rangle = 0$,
where the average kinetic energy is $\langle T \rangle = (d/2) N k_B
T$, and the full virial ${\cal W}$ is $W$ plus the contributions from
the forces from both domains walls at $R_a$ and $R_b$. However, the scaling
argument presented above is more general and can be adapted to other
problems where the virial theorem does not apply, for example in non
bounded systems, such as the cylindrical geometry presented below.

\subsubsection{Cylindrical geometry ($d=3$)}

An intermediate case between 2D and 3D is the cylindrical geometry, where
$Q_a$ and $Q_b$ are the charges of two concentric cylinders with
radius $R_a$ and $R_b$, and length $L\to\infty$. The volumes become 
${\cal V}_{a} = \pi R_a^2 L$ and ${\cal V}_{b} = \pi R_b^2 L$.
The interaction
potential between the ions is the 3D Coulomb potential $v(r)=1/r$, but
the interaction between the inner cylinder and an ion is logarithmic
$v_a(r)=-2(Q_a/L) \ln r$. The counterpart of Eq.~(\ref{eq:cont1}) follows
from noticing that volume changes to the cell are conceived transversally
to the cylinder. Therefore, the contact theorem reads identical to
Eq.~(\ref{eq:cont1}) with $d=2$, and the virial $W$ defined as
\begin{equation}
  W/e^2=\sum_{i<j} q_{i}q_{j}\frac{{r_{ij}^{\perp}}^2}{r_{ij}^{3}} 
  +2 \frac{Q_a}{L} \sum_{i} q_i
\end{equation}
where $r_{ij}$ is the distance between the ion $i$ and the ion $j$,
and $r_{ij}^{\perp}$ is the norm of the projection on a transversal
plane to the cylinders of the position vector between ions $i$ and
$j$. The equivalent to Eq.~(\ref{eq:cont1}) reads, in the cylindrical geometry,
specializing to a one-component system (ion charge $q$),
\begin{equation}
  \label{eq:contact-cyl-3d}
  {\pi R^2_b} n(R_b)=
  \frac{N}{L} + \frac{e^2 Q_a q N}{L^2} +  \frac{\beta (qe)^2}{2 L}
  \left\langle \sum_{i<j}
  \frac{{r_{ij}^{\perp}}^2}{r_{ij}^{3}}  \right\rangle
  + \pi  R^2_a n(R_a)
  \,.  
\end{equation}

\section{Explicit expressions and the planar limit}
\label{sec:explicit_then_planar}

The previous contact-like relations, Eqs. \eqref{eq:cont1}  and \eqref{eq:contact-cyl-3d},
establish a connection between the contact densities $n(R_a)$, $n(R_b)$, the surface charges
$\sigma_a$, $\sigma_b$, and the mean value of some known function of ions' coordinates.
It is instructive to analyze separately the three different geometries depicted in 
Figs. \ref{fig:art_sph} and \ref{fig:cyl_artistic}, in order to simplify the end result
to the greatest extent. This is the goal of the present section, where in each 
case, it will be checked that upon taking the planar limit, one recovers  as expected
the constraint \eqref{eq:contact_planar}.

\subsection{Two dimensions}
\label{sec:2d}

Using the explicit expression~(\ref{eq:w2d}) of $W$ in two dimensions,
Eq.~(\ref{eq:cont1}) becomes
\begin{equation}
  \label{eq:cont2d}
  R_{a}^2 \left( \sum_{\alpha} n_{\alpha}(R_a) - \pi \beta
    e^2\sigma_a^2\right)
  +\frac{1}{\pi} \sum_{\alpha} N_{\alpha}
  \left(1-\frac{\beta e^2 q_{\alpha}^2}{4}\right)
  =  R_{b}^2 \left( \sum_{\alpha} n_{\alpha}(R_b) - \pi \beta
        e^2\sigma_b^2\right)
\end{equation}
Interestingly, if there is only one type of particles in the system,
this expression can be ``separated'' into terms depending only on each
boundary:
\begin{equation}
  \label{eq:cont2docp}
  R_{a}^2\left[n(R_a)-\pi\beta e^2\sigma_{a}^2
    + \frac{2 e\sigma_{a}}{q R_{a}}
    \left( \frac{\beta e^2 q^2}{4}-1 \right) 
  \right]
  =
   R_{b}^2\left[n(R_b)-\pi\beta e^2\sigma_{b}^2
    - \frac{2 e\sigma_{b}}{q R_{b}}
    \left( \frac{\beta e^2 q^2}{4}-1 \right) 
  \right]
  \,.
\end{equation}
This is obtained using the electro-neutrality condition $Nq= -2 \pi
(\sigma_{a} R_{a}+ \sigma_{b} R_{b})$. Note that the ratio $\sigma_i/q$
is negative ($i=a$ or $i=b$).

The contact theorem for planar walls can be recovered. In the limit
$R_{a}\to\infty$ and $R_{b}\to\infty$ with $R_{b}-R_{a}=h<\infty$, the
``curvature'' terms $\pm \frac{2\sigma}{q R} \left( \frac{\beta e^2
    q^2}{4}-1 \right)$ from~(\ref{eq:cont2docp}) vanish, and
    introducing the coupling parameter $\Gamma = \beta e^2$ we obtain
the well-known expression
\begin{equation}
  \label{eq:contplan2d}
  n(R_a)-\pi\Gamma \sigma_{a}^2 = 
  n(R_b)-\pi\Gamma \sigma_{b}^2 
  \,.
\end{equation}
This is the counterpart, for a 2D system, of the constraint
\eqref{eq:contact_planar} put forward in the Introduction.
For a multicomponent electrolyte, the term $\sum_{\alpha} N_{\alpha}
\left(1-\beta e^2 q_{\alpha}^2/4\right)$ should be extensive in
the planar limit, i.e.~proportional to $R_{a} h$. Therefore it is
negligible in front of the other terms of Eq.~(\ref{eq:cont2d}), which
are proportional to $R_{a}^2$ or $R_{b}^2$, and we recover
again~(\ref{eq:contplan2d}).

\subsection{Three dimensions -- spherical geometry}
\label{sec:3d}

In three dimensions, by using relation~(\ref{eq:W-H}) between the virial
and the Hamiltonian, Eq.~(\ref{eq:cont1}) can be written as,
\begin{equation}
  \label{eq:cont3d}
  R_{a}^3 \left(\sum_{\alpha} n_{\alpha}(R_a)-2\pi\beta e^2\sigma_{a}^2
  \right)
  +\frac{1}{4\pi} \left(3\sum_{\alpha} N_{\alpha} + \beta \langle H
    \rangle \right) =
  R_{b}^3 \left(\sum_{\alpha} n_{\alpha}(R_b)-2\pi\beta e^2 \sigma_{b}^2
  \right)
  \,.
\end{equation}
Contrary to the two dimensional case, this expression cannot be
``separated'' into contributions from each boundary, even in the case
of a single component system. Reintroducing the permittivity of the solvent that was omitted
in the Hamiltonian (i.e. substituting $\beta e^2$ by $\ell_B$), Eq.~(\ref{eq:cont3d})
simplifies to
\begin{equation}
  \label{eq:cont3docp}
  R_{a}^3 \left( n(R_a)-2\pi\ell_B\sigma_{a}^2
    -\frac{3\sigma_a}{qR_{a}}\right)
    +\frac{\beta \langle H \rangle }{4\pi} =
  R_{b}^3 \left(n(R_b)-2\pi\ell_B\sigma_{b}^2
    +\frac{3\sigma_b}{qR_{b}}\right)
  \,.
\end{equation}
In the planar limit, the term $3 N -\beta \langle H \rangle$ is
extensive, i.e.~proportional to $R^{2}_{a} h$, therefore it is
negligible in front of the other terms of Eq.~(\ref{eq:cont3d}) which
are proportional to $R_{a}^3$ or $R_{b}^3$. Then, we have
\begin{equation}
  \label{eq:cont3dplan}
  n(R_a)-2\pi\ell_B \sigma_{a}^2 = 
  n(R_b)-2\pi\ell_B \sigma_{b}^2 
  \,.
\end{equation}
and we recover the
contact theorem \eqref{eq:contact_planar} for planar interfaces, in three dimensions.
It can be noted that relation \eqref{eq:cont3dplan} does also apply 
in the mean-field Poisson-Boltzmann framework (which in itself is noteworthy), where it bears the name
of Grahame equation \cite{Grahame47}.

\subsection{Three dimensions -- cylindrical geometry}
\label{sec:gen-cyl}

To investigate the situation corresponding to
Fig. \ref{fig:cyl_artistic}, we introduce the two body correlation
function $n_{\alpha\gamma}^{(2)} $ between ions of charge $q_{\alpha}$
and $q_{\gamma}$. Let $z$ be the component of $\r_1-\r_2$ along the
axis of the cylinders, and $\r_{1,2}^{\perp}$ the transverse
components. For an infinite cylinder ($L\to\infty$) the correlation
function depends only on $z$, $\r_{1}^{\perp}$ and
$\r_{2}^{\perp}$. In terms of the total correlation function
$h_{\alpha\gamma}$, the correlation function is
$n_{\alpha\gamma}(\r_1,\r_2)=n_{\alpha}n_{\gamma}(1+h_{\alpha\gamma}(\r_{1}^{\perp},\r_{2}^{\perp},z))$,
where $n_{\alpha}$, $n_{\gamma}$ are the average densities of
particles of species $\alpha$ and $\gamma$. It is shown in Appendix
\ref{app:h_cyl} that
\begin{multline}
\label{eq:cont-cyl}
R_b^2 (n(R_b)-2\pi\ell_B \sigma_b^2)
-R_a^2 (n(R_a)-2\pi\ell_B \sigma_a^2)
= \\
\frac{N}{\pi L} +\frac{e^2}{2\pi}
  \int d^2\r_{1}^{\perp}d^2\r_{2}^{\perp}
  \int_{-\infty}^{+\infty} dz \, e^2\sum_{\alpha \gamma} q_{\alpha} q_{\gamma}\,
  n_{\alpha}n_{\gamma}h_{\alpha\gamma}(\r_1^{\perp},\r_2^{\perp},z)
  \frac{{r_{12}^{\perp}}^2}{\left({r_{12}^{\perp}}^2+z^2\right)^{3/2}}
\,.
\end{multline}
A few limiting cases can be obtained from here. At the mean field level, $h_{\alpha\gamma}=0$, then the previous relation reduces to
\begin{equation}
\label{eq:cont-cyl-meanfield}
R_b^2 (n(R_b)-2\pi\ell_B \sigma_b^2)
-R_a^2 (n(R_a)-2\pi\ell_B \sigma_a^2)
= 
\frac{N}{\pi L} 
\,.
\end{equation}
It is interesting to note that this relation can be straightforwardly recovered from Eq. 
\eqref{eq:cont2d}, specified to the mean-field limit in which $\beta e^2$ does vanish,
the different valencies $q_\alpha$ being fixed. Indeed, 
the mean-field limit is described by a partial differential equation (the Poisson-Boltzmann
framework \cite{Levin02}), and {\em does not depend on the dimension of the system}.
This means that a circular charged rim in a concentric Wigner-Seitz circle, leads to the same
electrostatic potential as a charged cylinder inside a concentric Wigner-Seitz cylinder.
This is quite remarkable since the starting Hamiltonians, before taking the 
limit of weak coupling, differ somewhat. We see here an illustration of this
property, since enforcing $\beta e^2 \to 0$ in \eqref{eq:cont2d}
yields \cite{rque75}
\begin{equation}
  R_{a}^2 \left( \sum_{\alpha} n_{\alpha}(R_a) - \pi \beta
    \sigma_a^2\right)
  +\frac{1}{\pi} \sum_{\alpha} N_{\alpha}
  =  R_{b}^2 \left( \sum_{\alpha} n_{\alpha}(R_b) - \pi \beta
        \sigma_b^2\right),
\end{equation}
which is the counterpart of \eqref{eq:cont-cyl-meanfield}
(the two dimensional and three dimensional cases are connected through the substitution 
$\Gamma = \beta e^2 \leftrightarrow 2\ell_B$ and $N \leftrightarrow N/L$).

Beyond mean-field, that is for general coupling, the planar limit, $R_a\to \infty$, $R_b\to
\infty$ with $h=R_b-R_a$ finite, is recovered by noticing that the
right hand side of~(\ref{eq:cont-cyl}) is of order $R_{a}$ (or $R_b$),
while the left hand side is of higher order $R_a^2$, then
\begin{equation}
  n(R_b)-n(R_a)=2\pi\ell_B(\sigma_b^2- \sigma_a^2)
\,.
\end{equation}
We expectedly recover the planar contact theorem, see e.g. \eqref{eq:cont3dplan},
or equivalently \eqref{eq:contact_planar}. More generally,
for a one-component system, Eq.  (\ref{eq:cont-cyl}) becomes
\begin{multline}
  \label{eq:cyl-ocp}
  R_b^2 \left(n(R_b)-2\pi\ell_B \sigma_b^2+\frac{2 e \sigma_b}{R_b q}\right)
  -R_a^2 \left(n(R_a)-2\pi\ell_B \sigma_a^2-\frac{2 e \sigma_a}{R_a q}\right)
  = \\
  \frac{n^2\ell_B\, q^2}{2\pi}
  \int d^2\r_{1}^{\perp}d^2\r_{2}^{\perp}
  \int_{-\infty}^{+\infty} dz \,
  h(\r_1^{\perp},\r_2^{\perp},z)
  \frac{{r_{12}^{\perp}}^2}{\left({r_{12}^{\perp}}^2+z^2\right)^{3/2}}
  \,.
\end{multline}

\section{Application I : cylindrical colloids}
\label{sec:applI}

Knowing the generalization of the contact relation
Eq.~(\ref{eq:contact_planar}) to curved geometries, we are in a
position to discuss several applications. First, we show how known
results can be readily recovered for the two dimensional case. Then,
new results will be derived for the screening of three dimensional
cylinders, where obtaining the contact densities in closed form is not
possible. Accurate analytical expressions will be derived, and a
by-product of the analysis will be an expression for the fraction $f$
of condensed ions at finite density, whereas the celebrated Manning
scenario \cite{Manning} prescribes $f$ at infinite dilution only (where it takes the
value $f_M = 1-1/\xi$, $\xi$ being the dimensionless line charge to be
defined below).


\subsection{Screening of a two dimensional disk}

We consider the 2D case, with a one-component system of
counterions.  Let $\Gamma=\beta e^2$ be the coulombic coupling
constant. For $\sigma_b=0$ and by electro-neutrality, the
total number of ions is $N=|Q_a/q|$. Eq.~(\ref{eq:cont2docp}) reads
\begin{equation}
  \label{eq:nRaRb2D}
  \pi R_a^2 n(R_a) - \frac{\Gamma}{4} \left(\frac{Q_a}{q}\right)^2
  -\left| \frac{Q_a}{q}\right| \left(\frac{\Gamma}{4}-1\right) = \pi
  R_b^2 n(R_b) \,.
\end{equation}
Now, let us investigate the situation when $R_b\to\infty$. If
$R_b=\infty$, the derivation presented in section \ref{sec:sum-rule}
can be adapted. However, this system presents the
Manning condensation phenomenon \cite{rque50}, where only a partial fraction of the
ions remain bound to the charged disk \cite{NN05_PRL,NN06_PRE,BO05}. If $R_b=\infty$, the partition
function $Z^*$ of Eq.~(\ref{eq:Zstar}) is not properly defined,
unless it is restricted only to the number of condensed ions, as 
unbound ions give divergent contributions. Thus, in (\ref{eq:Zstar}), $N$ should be replaced by
$N_c$ which is the number of condensed ions onto the disk. The analog
of (\ref{eq:nRaRb2D}) is 
\begin{equation}
  \label{eq:nRa-2D-gen}
  \pi R_a^2 n(R_a) = -N_c
  \left[1-\frac{\Gamma}{4}
    + \frac{\Gamma}{4} \left(  N_c- 2\left| \frac{Q_a}{q}\right| \right)
    \right]
\end{equation}
Comparison with (\ref{eq:nRaRb2D}) yields the density at the outer
disk in the limit $R_b\to\infty$,
\begin{equation}
  \label{eq:nRb-2D-gen}
  \pi R_b^{2} n(R_b)= \left(\left|\frac{Q_a}{q}\right|-N_c\right)
  \left(
    1-\frac{\Gamma}{4}-\frac{\Gamma}{4}
    \left(\left|\frac{Q_a}{q}\right|-N_c\right)
  \right)\,.
\end{equation}
If $R_b$ is very large but not infinite, the picture of the separation
of the systems into two fluids, formed by the condensed counterions
and the unbound one holds, and equations (\ref{eq:nRa-2D-gen}) and
(\ref{eq:nRb-2D-gen}) should be valid. The condensed number of
counterions is \cite{NN06_PRE,BO05,JPMThesis13,New1}
\begin{equation}
  \label{eq:Nc2D}
  N_c=\ceiling{\left|\frac{Q_a}{q}\right| - \frac{2}{\Gamma}}
  =\left|\frac{Q_a}{q}\right| - \floor{\frac{2}{\Gamma}}
\end{equation}
where $\ceiling{x}$ and $\floor{x}$ are the ceiling and floor
functions. {The last equality in \eqref{eq:Nc2D} is only valid when
$|Q_a/q|$ is an integer, which is case here since $|Q_a/q|=N$.}
 \emma{It should be kept in mind that $N_c$ should remain positive,
which is not always the case with formula \eqref{eq:Nc2D}. It is therefore
understood that whenever \eqref{eq:Nc2D} leads to a negative quantity
($\floor{2/\Gamma} > |Q_a/q|$), 
$N_c=0$, meaning that counter-ion evaporation is complete}.
Here the number of condensed ions $N_c$ was
obtained as follows. It is the smallest number of counterions such
that the partition function of the disk of charge $Q_a$ with $N_c$
condensed counterions plus one additional unbound counterion is
divergent when $R_b\to\infty$, indicating that the additional
counterion is really unbound from the disk. According to this
definition, if the system has $N_c-1$ counterions bound to the disk,
it is able to bind one last additional charge \cite{rque71}. 

Replacing~(\ref{eq:Nc2D}) into (\ref{eq:nRa-2D-gen}) gives the density
at contact with the charged disk 
\begin{equation}
  \label{eq:nRa2Dfinal}
  \pi R_a^2 n(R_a) = 
  \frac{\Gamma}{4} \left(\frac{Q_a}{q}\right)^2
  +\left| \frac{Q_a}{q}\right| \left(\frac{\Gamma}{4}-1\right) 
  +\floor{\frac{2}{\Gamma}}
  \left(1-\frac{\Gamma}{4}-\frac{\Gamma}{4}\floor{\frac{2}{\Gamma}}
    \right)
    \,,
\end{equation}
\emma{with correspondingly,}
\begin{equation}
  \label{eq:nRb2D}
  \pi R_b^2 n(R_b) = 
  \floor{\frac{2}{\Gamma}}
  \left(1-\frac{\Gamma}{4}-\frac{\Gamma}{4}\floor{\frac{2}{\Gamma}}
    \right)
    \,,
\end{equation}
\emma{The above expressions hold provided evaporation is not complete,
while for $\floor{2/\Gamma}> |Q_a/q|=N$, we have $n(R_a)=0$ 
and
\begin{equation}
\pi R_b^2 n(R_b) = - \frac{\Gamma}{4} \left(\frac{Q_a}{q}\right)^2
  -\left| \frac{Q_a}{q}\right| \left(\frac{\Gamma}{4}-1\right).
  \label{eq:nRb2Dbis}
\end{equation}
}

\emma{These results deserve several comments. 
At arbitrary coupling, the planar limit 
\eqref{eq:contplan2d} should be recovered for $R_a\to \infty$, and fixed $\sigma_a$
(with thus $Q_a\to\infty$). This is indeed the case, since
linear terms in $Q_a$ can be neglected against quadratic ones in \eqref{eq:nRa2Dfinal}.
Second, they reproduce the mean-field limit, 
as it should, for $\Gamma \to 0$. 
This can be checked enforcing condensation to occur ($\floor{2/\Gamma}<N$).
To ensure compatibility of this constraint with the limit $\Gamma \to 0$,
we can work at fixed $\Gamma N$ and $N\to \infty$.
Equation \eqref{eq:nRa2Dfinal} then yields,
neglecting a term in $N \Gamma$ against those in $1/\Gamma \propto N$ 
\begin{equation}
\widetilde \rho(R_a) \,=\, \frac{n(R_a)}{\pi \Gamma \sigma_a^2} \, \simeq \, \frac{1}{4 \pi^2 R_a^2 \sigma_a^2\Gamma} \left(
\Gamma N^2 - 4 N + \frac{4}{\Gamma}
\right)  \, = \, 
\left(\frac{N\Gamma-2}{N\Gamma}\right)^2.
\end{equation}
With the substitution $N\Gamma \to 2\xi$, this is precisely of the Poisson-Boltzmann form 
\eqref{eq:rhocMFcyl}, } 
{valid when $\xi>1$.} {Turning to the contact density at $R_b$, we get from 
\eqref{eq:nRb2D} that 
\begin{equation}
  \label{eq:nRb2D-PB}
   \pi R_b^2 n(R_b) = \frac{1}{\Gamma}\,,
   \qquad \Gamma\to 0\,.
\end{equation}
which indeed is the mean-field expression, reminded in (\ref{eq:nRb-PB}) below}.
{
For the particular case of $2/\Gamma>N$ (i.e. $\xi<1$) we obtained that $N_c=0$,
hence, the density at $R_a$ is trivial: $n(R_a)=0$. On the outer shell,
\emp
\pi R_{b}^2n(R_b)=N\Bpar{1-\frac{N\Gamma}{4}\Rpar{1+\frac{1}{N}}}.
\label{eq:rho2d_Rb_Nc0}
\fin
This is fully compatible with the result from Poisson--Boltzmann theory
\emp
\widetilde{\rho}(R_B)=\frac{\pi R_{a}^2n(R_b)}{\pi\Gamma\sigma_{a}^{2}}=\Rpar{\frac{R_a}{R_b}}^2\frac{2}{N\Gamma}\Rpar{2-\frac{N\Gamma}{2}}.
\label{eq:rho2d_Rb_Nc0_MF}
\fin
}
However, {\em for arbitrary $\Gamma$}, the result for $n(R_b)$ departs from mean-field,
which might come as a surprise since the charged fluid of counterions
becomes extremely dilute at $R_b$. In three dimensions, diluteness ensures
that mean-field applies far from the charged cylinder,
irrespective of the strength of coupling \cite{MTT13}. In 2D on the other hand, no matter how far from the charged
cylinder the counterions are, they are still coupled, due to the scale
invariance of the logarithmic interaction \cite{Varenna}. 

\emma{Third}, for $\Gamma\geq 2$, the right hand side of (\ref{eq:nRb2D})
vanishes, showing that $n(R_b)$ decays faster than $R_b^{-2}$. This can be understood by noticing that for
$\Gamma\geq 2$ the number of condensed counterions is
$N_c=|Q_a/q|$. Thus, far from the disk, the effective potential that
one single ion of charge $q$ at a distance $r$ feels, is that of the
charged disk plus the $N-1$ remaining condensed ions. That object
has a total charge $-q$, which therefore leads to an effective potential of the form
$U_{\text{eff}}(r) = q \ln r$. One should consequently expect that the density
behaves as $n(r)\sim e^{-\beta q U_{\text{eff}}(r)} = r^{-\Gamma}$ and
it does decays faster that $r^{-2}$ when $\Gamma\geq 2$. For
$\Gamma=2$, an exact result \cite{JPMThesis13,New1} shows that $R_b^{2} n(R_b) \sim
1/(2\ln (R_b/R_a))$ as $R_b\to\infty$.
Finally, we show in Fig. \ref{fig:2D_contact_D100} that Eqs.~(\ref{eq:nRa2Dfinal}) and (\ref{eq:nRb2D})
are in excellent agreement with the Monte Carlo data, provided $R_b$ is large enough. 
The notation for the densities used in the plots corresponds to a rescaling with the exact planar  result; i.e. $\widetilde{\rho}\defeq\,n/n_{\text{plate}}\equiv\,n/\Rpar{\pi\Gamma\sigma_{a}^{2}}$. 
The figure corresponds to $R_b/R_a=e^{100}$, while  decreasing this ratio 
leads to rather strong finite size effects, that will be
studied elsewhere \cite{New1}.
\emma{We note that for $\Gamma<2/N$, $n(R_a)=0$, as a fingerprint of the vanishing of $N_c$
(to anticipate a coming and often used notation, this corresponds to $\xi<1$ \cite{rque50}).}

\psfrag{AX}{$\Gamma$}
\psfrag{BY}{{\color{OliveGreen} $\widetilde{\rho}\vert_{r=R_a}$}}
\psfrag{BY2}{{\color{red} $(N\Gamma/2)^2(R_{b}/R_{a})^{2}\widetilde{\rho}\vert_{r=R_b}$}}
\ImT{0.5}{contact_D100}{2D_contact_D100}{The symbols show the density $\widetilde{\rho}$ at contact in $r=R_a$ (green) and $r=R_b$ (red) 
as a function of the coupling parameter $\Gamma$, as obtained in Monte Carlo simulations; 
here, $\log (R_b/R_a)=100$, \emma{$N=|Q_a/q|=10$ and $q=1$}. The tilde notation follows
from rescaling the densities with the value in the planar case $\pi \, \Gamma\sigma_{a}^{2}$ such that
$
\widetilde{\rho} = n/(\pi\Gamma\sigma_{a}^2).
$
\emma{The dashed curves represent the $R_b\to\infty$ formulation from
\cref{eq:nRa2Dfinal} supplemented with $n_a=0$ for $\Gamma <2/N$, and \cref{eq:nRb2D,eq:nRb2Dbis} as far as the density at the outer boundary $R_b$ is concerned. 
The arrows indicate the location of
$\Gamma=2$, the borderline to complete condensation beyond which $n(R_b)=0$}.}

In the strong coupling limit \emma{(large $\Gamma $)} and in absence of an external boundary charge,
  condensation is complete: the number of condensed ions is maximal,
  $N_c=N$. \emma{More precisely, this occurs as soon as $\Gamma>2$.}
  For such a situation, Eq.~(\ref{eq:nRa2Dfinal}) becomes 
\emp
\label{eq:2d_sc_contacr_R}
\pi R_{a}^{2}n\Rpar{R_a}\, =\,
\frac{N^2\Gamma}{4}\Rpar{1-\frac{4}{N\,\Gamma}+\frac{1}{N}}\,.  \fin
It can be seen in \cref{fig:contactSC_2d} that this expression 
coincides with the Monte Carlo measures, for $\Gamma >2$. Note that
\eqref{eq:2d_sc_contacr_R} carries the leading order from the planar
limit, \emph{i.e.} $\pi R_{a}^{2}n\Rpar{R_a}=N^2\Gamma/4$. The
result~(\ref{eq:2d_sc_contacr_R}) can indeed be recovered by
adapting the Wigner strong coupling approach presented in Refs~\cite{WSC1,
  WSC2} to the present case. Implementing this technique turns
out to slightly differ from the planar
geometry (corresponding to a line in 2D) where the profile, to leading
order, is given by the interaction with the surface charge alone. Here,
the remaining condensed counterions also contribute
to the profile to leading order; thus, the contact density carries
this trait as well. The details of the derivation are presented
elsewhere~\cite{JPMThesis13, New1} with the result that the density profile
behaves as $n(r)\sim\,r^{-\frac{N\Gamma}{2}-\frac{\Gamma}{2}}$ and the
corresponding density at contact is precisely given
by~(\ref{eq:2d_sc_contacr_R}).

\psfrag{AX}{$\Gamma$}
\psfrag{BY}{$\widetilde{\rho}\vert_{r=R_a}$}
\ImT{0.5}{contactSC_2d}{contactSC_2d}{The symbols show the density $\widetilde{\rho}$ at contact ($r=R_a$), as a function of the coupling parameter $\Gamma$, for 
$N=5$ and $N=10$. The dashed and dashed-pointed curves represent the analytic prediction for full condensation from \cref{eq:2d_sc_contacr_R}. The dotted arrow pointing upwards shows the threshold to full condensation.}

\subsection{Screening of a cylindrical macro-ion}
\label{ssec:cyl}

In this section, we focus on the cylindrical geometry (see Fig. \ref{fig:cyl_artistic}), where only the
inner cylinder is charged ($Q_b=0$), and  is screened  by
counterions of charge $q$. To make the connection with
previous works~\cite{NN05_PRL}, it is convenient to introduce the
following notations: the Manning
parameter $\xi=2\pi \ell_B R_a q |\sigma_a|$, the Coulomb coupling
parameter $\Xi=2\pi \ell_B^2 q^3 |\sigma_a|$ \cite{Netz01,WSC1,Varenna}, and
$\widetilde{\rho}(r)=n(r)/(2\pi\ell_B \sigma_a^2)$. Due to
electro-neutrality, the total number of counterions is such that $N/L=\xi/\ell_B$,
with a slight abuse of language (we deal with systems of infinite length $L$,
with thus a divergent $N$). 
With these notations, the relation (\ref{eq:contact-cyl-3d}) reads
\begin{equation}
  \label{eq:cont-cyl-ocp}
  \widetilde{\rho}(R_a)=2\frac{\xi-1}{\xi}+\left(\frac{R_b}{R_a}\right)^2 
  \widetilde{\rho}(R_b)-\frac{\ell_B^2}{2\xi^2 L}
  \left\langle \sum_{i}\sum_{j\neq i} \frac{{r_{ij}^{\perp}}^2}{r_{ij}^3}
  \right\rangle
\end{equation}
Next, we consider the situation when $R_b\to\infty$. If
$R_b=\infty$, the derivation presented in section \ref{sec:sum-rule}
should be adapted. In the cylindrical geometry, again, only a partial
fraction $f$ of the ions remain bound to the charged cylinder if $R_b$
is very large. Thus, in (\ref{eq:Zstar}), $N$ should be replaced by
$N_c=f N$ the number of condensed counterions, where $f$ is the fraction 
of such ions. Then, when
$R_b\to\infty$, Eq.~(\ref{eq:cont1}) 
 \begin{equation}
  \label{eq:cont3}
  L \pi R_a^2 n(R_a) =
  -N_c +  \frac{1}{2} \langle \beta W \rangle
  \,.
\end{equation}
where as above, it should be understood that we consider the limit $L\to\infty$.
In $W$, only the contribution from the condensed ions should be included.
That is
\begin{equation}
  \label{eq:cont-cyl-ocp-Rbinfty-general-f}
  \widetilde{\rho}(R_a)=2\left(\frac{\xi-1}{\xi}\right) f 
  -\frac{\ell_B^2}{2\xi^2 L}
  \left\langle \sum_{i\in \mathcal{B}}\sum_{j\in \mathcal{B}, j\neq i} \frac{{r_{ij}^{\perp}}^2}{r_{ij}^3}
  \right\rangle
\end{equation}
where $\mathcal{B}$ is the set of bound (condensed) ions to the charged
cylinder. Comparing (\ref{eq:cont-cyl-ocp}) to
(\ref{eq:cont-cyl-ocp-Rbinfty-general-f}), one can deduce that in the limit
$R_b\to\infty$, the density at the outer cell, $n(R_b)$, satisfies
\begin{equation}
  \label{eq:nRb-general-f}
  \left(\frac{R_b}{R_a}\right)^2
  \widetilde{\rho}(R_b)=
  2\frac{\xi-1}{\xi}(f-1)
  +\frac{\ell_B^2}{2\xi^2 L}\left[
    \left\langle \sum_{i}\sum_{j\neq i} \frac{{r_{ij}^{\perp}}^2}{r_{ij}^3}
  \right\rangle -
    \left\langle \sum_{i\in \mathcal{B}}\sum_{j\in \mathcal{B}, j\neq i} \frac{{r_{ij}^{\perp}}^2}{r_{ij}^3}
  \right\rangle
  \right]
\end{equation}
where in the averages, the first sum includes correlation between all
the counterions (both bound and unbound), whereas in the second sum
only correlations between the bound ions are taken into account.

\begin{figure}
  \centering
  \subfloat[][Needle]{%
  \includegraphics[height=4.5cm]{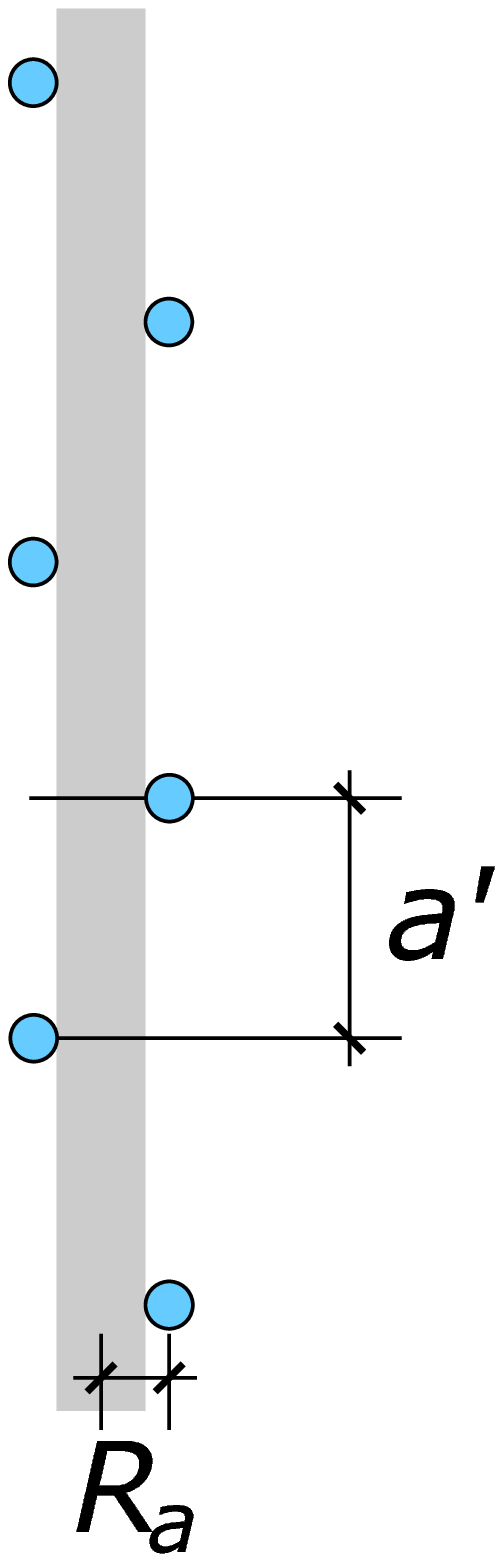}
  }%
  \quad\quad\quad\quad%
  \subfloat[][Diagram]{%
  \includegraphics[height=5.5cm]{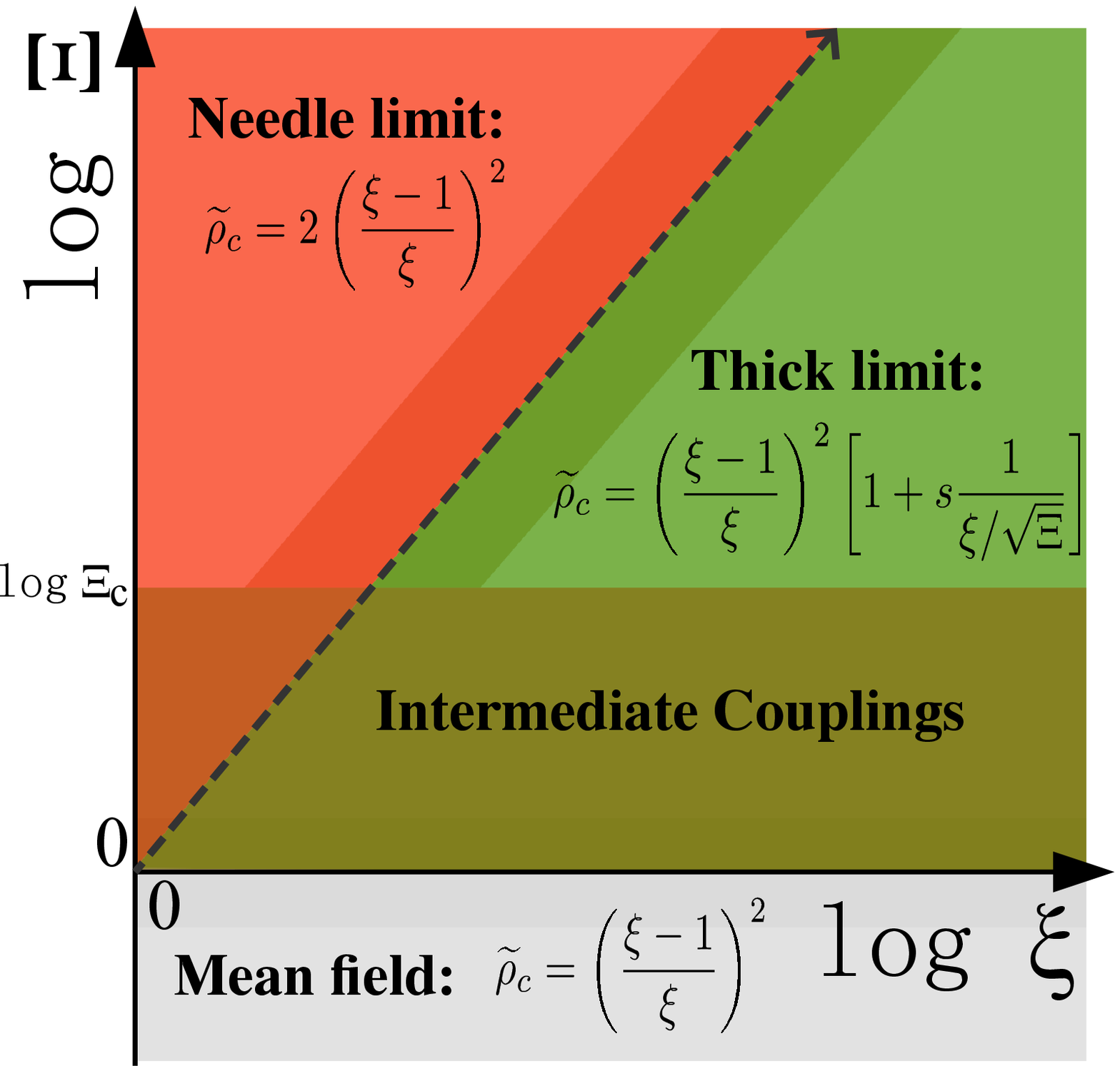}%
  \label{fig:phase_diag}}%
  \quad\quad\quad%
  \subfloat[][Thick]{%
  \includegraphics[height=4.5cm]{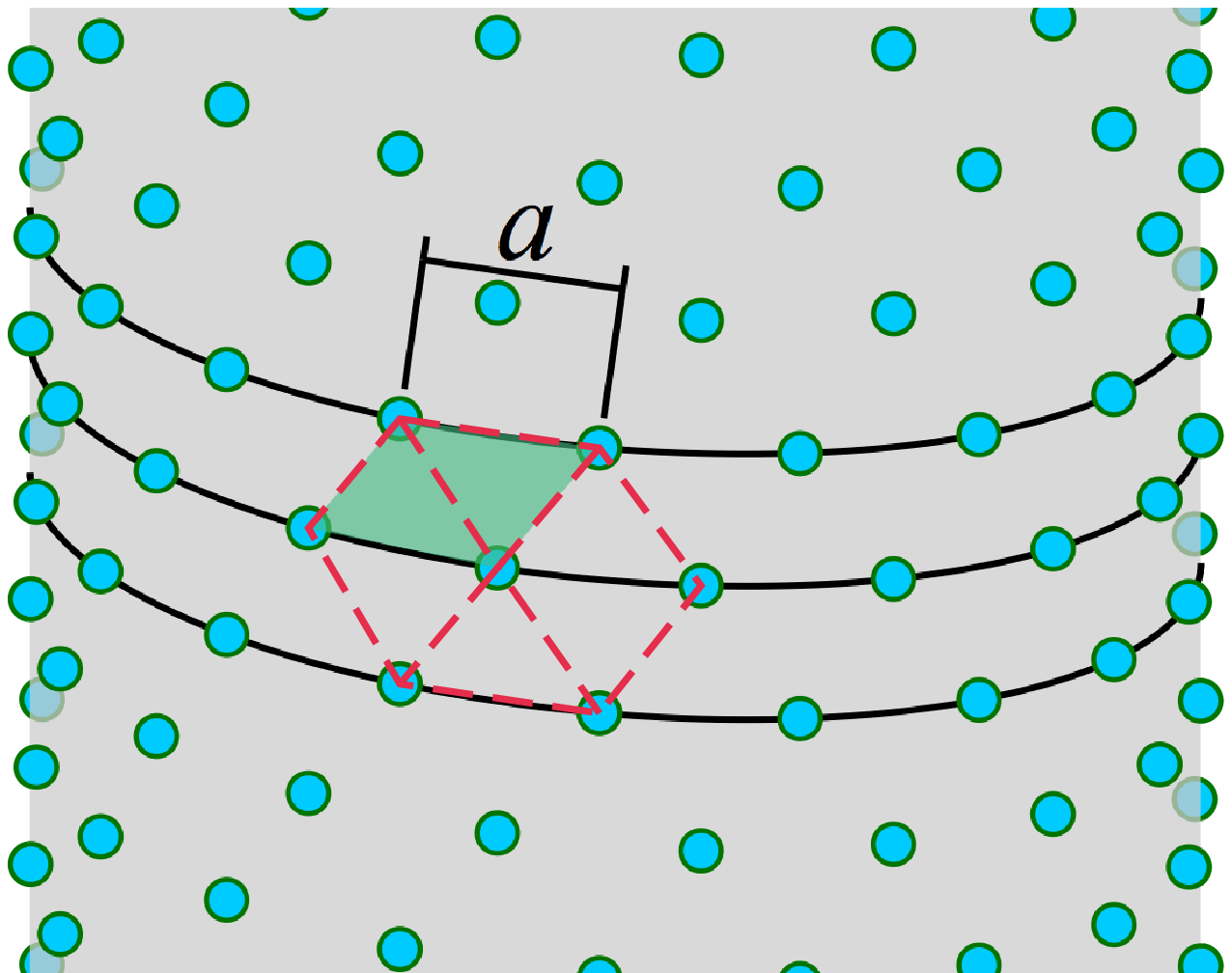}
  }%
  \caption{Summary diagram emphasizing the different regimes for the cylinder problem, and the value for the contact densities. 
  $(a)$ and $(c)$ represent 'artistic' views of the needle (thin) and thick limits. \emma{Both correspond to strong coupling $\Xi \gg 1$,
  with $\xi \ll \Xi^{1/2}$ (needle/thin, where ions form a quasi 1D Wigner crystal) or $\xi \gg \Xi^{1/2}$ (thick case, where ions form a curved
  2D crystal).}
  In panel $(b)$, the dashed line represents $\xi=\sqrt{\Xi}$, the borderline between the needle and the thick cylinder regimes. Poisson-Boltzmann theory applies, roughly speaking, for $\Xi<1$ while the upper part of the diagram is for the strong coupling regime $\Xi \gg1$. An intermediate region stands between the strong coupling needle and thick limits and mean field where the properties of the system are of neither nature.}
\label{fig:phase_diag2}
\end{figure}

If $R_b=\infty$, the fraction of condensed ions is
$f=f_M=(\xi-1)/\xi$. This is the celebrated Manning result \cite{DeHo01,Manning,TT06}. Then 
\begin{equation}
  \label{eq:cont-cyl-ocp-Rbinfty}
  \widetilde{\rho}(R_a)=2\left(\frac{\xi-1}{\xi}\right)^2
  -\frac{\ell_B^2}{2\xi^2 L}
  \left\langle \sum_{i\in \mathcal{B}}\sum_{j\in \mathcal{B}, j\neq i} \frac{{r_{ij}^{\perp}}^2}{r_{ij}^3}
  \right\rangle
\end{equation}
and
\begin{equation}
  \label{eq:nRb}
  \left(\frac{R_b}{R_a}\right)^2
  \widetilde{\rho}(R_b)=
  2\frac{1-\xi}{\xi^2}
  +\frac{\ell_B^2}{2\xi^2 L}\left[
    \left\langle \sum_{i}\sum_{j\neq i} \frac{{r_{ij}^{\perp}}^2}{r_{ij}^3}
  \right\rangle -
    \left\langle \sum_{i\in \mathcal{B}}\sum_{j\in \mathcal{B}, j\neq i} \frac{{r_{ij}^{\perp}}^2}{r_{ij}^3}
  \right\rangle
  \right]
\end{equation}
In the following, three different limit will be investigated. We will 
start by the mean-field Poisson-Boltzmann regime ($\Xi\ll1$) before taking the opposite 
view and work out the effects of strong correlations ($\Xi\gg1$). There, one should discriminate
the cases where $\xi \gg \Xi^{1/2}$ and $\xi \ll \Xi^{1/2}$, which can respectively
be coined ``thick'' and ``thin'' (or needle) situations \cite{MTT13}. Cases with $\xi \simeq \Xi^{1/2}$
correspond to a crossover where analytical progress is more difficult, and
will not be addressed. The difference between the thin and thick cases
can be appreciated pictorially in Fig. \ref{fig:phase_diag2}, panels a) and c).

\subsubsection{Mean field limit}

For $R_b\to\infty$, in the mean field approximation stemming from 
$\Xi\to 0$, the last term of
(\ref{eq:cont-cyl-ocp-Rbinfty}) can be computed following the same
lines as in Sec.~\ref{sec:gen-cyl}, neglecting the correlation
function $h$ and only considering the condensed number of ions
$N_c=(\xi-1) L/\ell_B$ instead of $N$. Then
\begin{equation}
\frac{\ell_B^2}{2\xi^2 L}
  \left\langle \sum_{i\in \mathcal{B}}\sum_{j\in \mathcal{B}, j\neq i} \frac{{r_{ij}^{\perp}}^2}{r_{ij}^3}
  \right\rangle  = 
  \left(\frac{\xi-1}{\xi}\right)^2
\end{equation}
and we obtain
\begin{equation}
  \label{eq:rhocMFcyl}
  \widetilde{\rho}(R_a) \,=\, \left(\frac{\xi-1}{\xi}\right)^2 \,=\, f_M^2 .
\end{equation}
We thereby recover the known value \cite{MTT13}, as following
from Poisson-Boltzmann theory, for an original approach.
The above expression assumes that $\xi>1$, so that counterion
condensation effectively takes place. All parameters being kept
fixed but the cylinder radius $R_a$, one should recover the planar
limit, reading here $\widetilde \rho_a = 1$, when $R_a \to\infty$.
Remembering that $\xi \propto R_a$,
this is indeed the case, as is seen by taking the limit $\xi\to\infty$
in Eq.~(\ref{eq:rhocMFcyl}).

\subsubsection{Strong coupling and large dilution -- Thin cylinder limit}

In the opposite limit (strong coupling regime) where $\Xi\gg 1$, an explicit calculation can be performed.
We furthermore need to assume the thin cylinder limit $R_a/a'=\xi^2 f/\Xi\ll 1$,
where $a'$ is the lattice constant of the 1D Wigner crystal formed by
the condensed ions along the cylinder~\cite{MTT13}, see Fig. \ref{fig:phase_diag2}a). If $\mathcal{U}$ denotes the
set of unbound ions, we have
\begin{equation}
  \label{eq:Wub}
    \left\langle \sum_{i}\sum_{j\neq i} \frac{{r_{ij}^{\perp}}^2}{r_{ij}^3}
  \right\rangle -
    \left\langle \sum_{i\in \mathcal{B}}\sum_{j\in \mathcal{B}, j\neq i} \frac{{r_{ij}^{\perp}}^2}{r_{ij}^3}
  \right\rangle
  = 2    \left\langle \sum_{i\in \mathcal{B}}\sum_{j\in \mathcal{U}} \frac{{r_{ij}^{\perp}}^2}{r_{ij}^3}
  \right\rangle +
    \left\langle \sum_{i\in \mathcal{U}}\sum_{j\in \mathcal{U}, j\neq i} \frac{{r_{ij}^{\perp}}^2}{r_{ij}^3}
  \right\rangle
  \,.
\end{equation}
Even if the coupling is strong near the charged cylinder, in the far
region, the unbound ions are diluted enough so that mean field applies to
them~\cite{MTT13}. Thus, the second term \emma{on the right hand side} of (\ref{eq:Wub}) can be
evaluated neglecting the correlations. Let $n_u(\r^{\perp})$ denote the density of uncondensed ions (which does not depend on the $z$ coordinate). Then, using (\ref{eq:useful-integral}) to perform the integrals along the axis of the cylinder,
\begin{eqnarray}
    \left\langle \sum_{i\in \mathcal{U}}\sum_{j\in \mathcal{U}, j\neq i} \frac{{r_{ij}^{\perp}}^2}{r_{ij}^3}
  \right\rangle
  &=& \int d^2\r_1^{\perp} d^2\r_2^{\perp}
  dz_1 dz_2 n_u(\r_1^{\perp}) n_u(\r_2^{\perp}) \frac{{r_{12}^{\perp}}^2}{r_{12}^3}
\nonumber  \\
  &=& 2 L \int d^2\r_1^{\perp} d^2\r_2^{\perp}
   n_u(\r_1^{\perp}) n_u(\r_2^{\perp})
   =2 N_u^2 / L
   \label{eq:inter1}
 \end{eqnarray}
 where $N_u=N-N_c$ is the total number of uncondensed ions. The other contribution to (\ref{eq:Wub}) can
 be computed by supposing that the uncondensed ions are completely
 uncorrelated from the condensed ones and form a gas of density
 $n_u(\r^{\perp})$
\begin{eqnarray}
  \left\langle \sum_{i\in \mathcal{U}}\sum_{j\in \mathcal{B}} \frac{{r_{ij}^{\perp}}^2}{r_{ij}^3}
  \right\rangle
  &=&   N_c  \left\langle \sum_{i\in \mathcal{U}} \frac{{r_{i}^{\perp}}^2}{r_{i}^3}
  \right\rangle 
\nonumber  \\
  &=& N_c
\int d^2\r^{\perp}  dz \, n_u(\r^{\perp})  \frac{{r^{\perp}}^2}{r^3}
\nonumber  \\
&=& 2 N_c \int d^2\r^{\perp}\,  n_u(\r^{\perp})  
\nonumber  \\
&=& 2 N_c N_u /L\,.
\label{eq:inter2}
\end{eqnarray}
So far, relations \eqref{eq:inter1} and \eqref{eq:inter2} hold, irrespective
of the value of the condensed fraction $f$, with $N_c = fN$ and $N_u = (1-f)N$.
They will therefore be used in the subsequent analysis, where because of
finite size effects, $f$ takes a non trivial value \emma{(and thus differs from $f_M=1-1/\xi$)}. Here, we consider the
case of large $\log(R_b/R_a)$, where $f\to 1-1/\xi$. In other words, 
we have now that $N_u/L=1/\ell_B$, $N_c/L = (\xi-1)/\ell_B$. 
Gathering results in (\ref{eq:nRb}), we get
\begin{equation}
  \label{eq:nRb-PB}
  \left(\frac{R_b}{R_a}\right)^2
  \widetilde{\rho}(R_b)=\frac{1}{\xi^2}
\end{equation}
which is consistent with the value given explicitly by the
Poisson-Boltzmann mean field solution~\cite{NN06_PRE,Fuoss01091951}.
Indeed, irrespective of the coupling parameter $\Xi$,
the ions far from the charged cylinder at $R_a$ are dilute enough
so that mean-field does hold.

This is not the case in the vicinity of the charged cylinder.
The density at contact
(\ref{eq:cont-cyl-ocp-Rbinfty}) can be
evaluated as follows. In the strong coupling and the thin cylinder limit,
the $z$ coordinates of condensed ions are fixed $z_n= n a'$ where $n$
is an integer (the ions form a quasi one dimensional Wigner crystal). The leading order contribution to the potential energy
of the system is given by the cylinder-ion terms, allowing for small
vibrations in the radial direction $\r^{\perp}$, with a one-body
Boltzmann factor proportional to ${r^{\perp}}^{2\xi}$. Thus
\begin{eqnarray}
  \left\langle \sum_{i\in \mathcal{B}}\sum_{j\in \mathcal{B}, j\neq i} \frac{{r_{ij}^{\perp}}^2}{r_{ij}^3}
  \right\rangle
  &=& N_c  \sum_{n\in\mathbb{Z}^{*}} \frac{2}{|n|^{3} a'^{3}} 
  \frac{  \displaystyle
    \int_{r^{\perp}>R_a} {r^{\perp}}^{2\xi}{r^{\perp}}^2 
    d^{2}\r^{\perp} }{   \displaystyle \int_{r^{\perp}>R_a} 
    {r^{\perp}}{^{2\xi}}  d^{2}\r^{\perp} } 
  \nonumber\\
  &=& 4 \zeta(3) f^3 \frac{\xi^5}{\Xi^2 \ell_B} N_c \frac{\xi-1}{\xi-2}
  \,,
\end{eqnarray}
where $\zeta(t)=\sum_{n=1}^{\infty} n^{-t}$ is the Riemman zeta
function. Then, replacing into (\ref{eq:cont-cyl-ocp-Rbinfty}), we have
\begin{equation}
  \label{eq:rhoRa}
  \widetilde{\rho}(R_a) = 2 \left(\frac{\xi-1}{\xi}\right)^2
  \left(1 -
    \left(\frac{\xi-1}{\xi}\right)^3 \frac{\xi}{\xi-2} 
    \left(\frac{\xi^2}{\Xi} \right)^2 \zeta(3)
    \right)
\end{equation}
This is in full agreement with the prediction from~\cite{MTT13}, where
it was obtained by completely different means, generalizing the route
outlined in \cite{WSC1,WSC2}. The term with $\zeta(3)$ should be seen
as a small correction, meaning that to leading order, the contact
density is twice its mean-field counterpart $(\xi-1)^2/\xi^2$, see
Eq.(\ref{eq:rhocMFcyl}).

\subsubsection{Strong coupling and thin cylinder limit: the contact density for finite $R_b$}

We have seen that under large dilution ($R_b\to\infty$), the fraction $f$
of condensed ions goes to its known Manning limit $f_M=1-1/\xi$. The same limiting expression
is reached in the mean-field regime as well ($\Xi\to 0$, at fixed $\xi$).
At arbitrary coupling $\Xi$ and at finite although 
large $R_b$, the situation is more complex, and plagued by severe finite
size effects \cite{NN05_PRL,NN06_PRE,MTT13}. It has been reported
that one has in general $f>f_M$, but no analytical 
expression is available in general for $f$. Under strong coupling $\Xi \gg 1$, an empirical equation
was put forward in \cite{MTT13}, which relates $f$ to
the coupling parameter $\Xi$ and $\log R_b/R_a$: 
\emp
\frac{f-f_M}{f_M}\simeq\frac{\log\Xi-\delta}{\log\frac{R_b}{R_a}}.
\label{eq:f_empirical}
\fin
For the range of values of $\xi$ (between 3 and 5) and $R_b/R_a$
explored in~\cite{MTT13}, $\delta\approx4.5$ within a
margin of 20\%. We will soon be in a position to provide a justification
of expression \eqref{eq:f_empirical} (section \ref{ssec:determine_f}), but before turning to these considerations,
we leave $f$ as an unknown parameter and check for the 
consistency of the contact relations in which it appears.

We assume that $R_b/R_a$ is finite, but large enough to allow for
a clear-cut distinction between bound and unbound populations. 
Going back to \eqref{eq:cont-cyl-ocp-Rbinfty-general-f} and neglecting the term in brackets,
which is valid under strong coupling, we have 
\eemp
\widetilde{\rho}(R_a)&\simeq 2 \,f_M\, f.
\label{eq:rhoRa_f}
\ffin
The term neglected on the r.h.s of \eqref{eq:cont-cyl-ocp-Rbinfty-general-f}, 
can be estimated to behave like  $\ell_B^2 N_c R_a^2 / (\xi^2 L a'^3) \propto (R_a^2/a'^2) \ll 1$,
from the needle constraint (see Fig. \ref{fig:phase_diag2}a), which can also be expressed  as $\xi \ll \sqrt{\Xi}$).
This very feature can also be observed in Eq. \eqref{eq:rhoRa}, where the 
term in $\zeta(3)$ is a correction to the dominant behavior.
Introducing a parameter $t\defeq\xi(f-f_M)$ ($t\in\Bpar{0,1}$), the contact density 
\eqref{eq:rhoRa_f} can be written as
\emp
\frac{\widetilde{\rho}(R_a)}{2f_{M}^2}\simeq{1+\frac{t}{\xi-1}}.
\label{eq:rhoRa_t}
\fin
This is in excellent agreement
with Monte Carlo data, see Fig.~\ref{fig:size_contact_R}, where 
$f$ is in general unknown, but measured from the measured profiles
following the inflection point criterion
often used to quantify condensation \cite{Belloni98,TT06,MTT13},
and to separate the bound from the unbound ions.

\psfrag{AX}{$t$}
\psfrag{BY}{$\widetilde{\rho}(R_a)/2f_{M}^{2}$}
\ImT{0.5}{contact_size_R_v2}{size_contact_R}{Contact density as a function of $t=\xi(f-f_M)$, which measures the deviation from
infinite dilution condensation fraction. The prediction from Eq.~(\ref{eq:rhoRa_t}) corresponds to the dashed lines in the plot.
\emma{Upon changing the system size (in the range $10 < \log(R_b/R_a))<100$), the fraction of condensed ions $f$ changes,
which causes $t$ to vary.}
Error bars are smaller than the tick size. 
}

\psfrag{AX}{$t$}
\psfrag{BY}{$\xi^2\Rpar{{R_b}/{R_a}}^2\, \widetilde{\rho}(R_b)$}
\psfrag{LLLLLLL1}{\fontsize{9}{0}\selectfont {\, \color{black} $\blacktriangle\, \, \blacksquare$ \, $\xi=3$}}
\psfrag{LLLLLLL2}{\fontsize{9}{0}\selectfont {\, \color{red} $\blacktriangle\, \, \blacksquare$} \, $\xi=4$}
\psfrag{LLLLLLL3}{\fontsize{9}{0}\selectfont {\, \color{OliveGreen} $\blacktriangle\, \, \blacksquare$} \, $\xi=5$}
\ImT{0.5}{contact_size_D_v2}{size_contact_D}{Same as Eq.~(\ref{fig:size_contact_R}) for the rescaled density at the outer surface $R_b$. 
Shown is $\xi^2(R_b/R_a)^2\, \widetilde{\rho}(R_b)$, as a function of $t$. Note that the analytic prediction from Eq.~(\ref{eq:rhoRb_t}), represented by the dotted curve, is independent of $\xi$.}

Conversely, for the outer shell, we have to proceed from Eq.~(\ref{eq:nRb-general-f}), where it is no longer
possible to neglect the terms in square brackets. Making use of Eqs. \eqref{eq:Wub}, \eqref{eq:inter1} and 
\eqref{eq:inter2}, we obtain 
\eemp
\Rpar{\frac{R_b}{R_a}}^2\widetilde{\rho}(R_b)=&(1-f_M)^2-(f-f_M)^2.
\label{eq:rhoRb_f}
\ffin
In $t$ representation, this gives
\eemp
\xi^2\Rpar{\frac{R_b}{R_a}}^2\, \widetilde{\rho}(R_b)=&1-t^2,
\label{eq:rhoRb_t}
\ffin 
which is a universal parabola, irrespective of $\xi$. Here also,
the comparison with Monte Carlo is very good, see
Fig.~\ref{fig:size_contact_D}.

\subsubsection{Estimation of the condensed fraction for finite $R_b$}
\label{ssec:determine_f}

We have so far derived contact relations at $R_a$ (inner surface)
and at $R_b$ (outer surface), which have been shown to be confirmed
by Monte Carlo simulations. These relations involve the condensed
fraction, $f$, which is only known in the truly dilute limit
where $R_b\to\infty$, in which case $f\to f_M = 1-1/\xi$.
However, $f$, as several other quantities of interest,
approaches its dilute limit in a logarithmic fashion,
and our goal here is to derive an explicit expression
for the size dependence of $f$. To this end, we develop in Appendix
\ref{APP_towards_f} a heuristic approach, which relies on the assumption
that the bound (condensed) and unbound fluids can be clearly separated.
The idea is to start from the following contact balance equation
\eemp
\Rpar{\frac{R_b}{R_a}}^2\frac{\widetilde{\rho}(R_b)}{\widetilde{\rho}(R_a)}=\frac{(1-f_M)^2-(f-f_M)^2}{2f_Mf},
\label{eq:contact_balEQ}
\ffin 
and to search for an alternative expression for the ratio of
densities appearing on the left hand side. 
This is
  done in Appendix \ref{APP_towards_f}. Combining \cref{eq:contact_balEQ} with
  \cref{eq:ratio-altern}, and making again use of $t = \xi(f-f_M)$, we
  obtain 
  \begin{align}
-\frac{\log\left(1-t^2\right)}{2\xi}+\Rpar{\frac{1+f_M}{2}+\frac{t}{\xi}}\log\Rpar{1+\frac{t}{f_M\xi}}+\frac{t}{\xi}\Kpar{\log\frac{R_b}{R_a}-\log\Xi-\frac{1+f_M}{2f_M}-\mathcal{B}\Rpar{\xi}}
=f_M\Kpar{\log\Xi-\mathcal{A}\Rpar{\xi,\frac{R_b}{R_a}}},
\label{eq:heuristic_f}
  \end{align}
with $\mathcal{A}$ and $\mathcal{B}$ given by,
\eemp
\mathcal{A}\Rpar{\xi,\frac{R_b}{R_a}}&=2\log\xi+\gamma-\log2+\log f_M+\frac{2\Rpar{\log\Bpar{\log\frac{R_b}{R_a}}-1}+\log\Bpar{2\xi^2f_{M}^2}-1}{2f_M\xi}\\
\mathcal{B}\Rpar{\xi}&=-2\log\xi-\gamma+\log2-\log f_M-\frac{1+f_M}{2f_M}.
\ffin

Equation~(\ref{eq:heuristic_f}) simplifies in the limit of small $t$ (i.e. large box size), where 
\eemp
t\simeq&f_M\xi\frac{\log\Xi-\mathcal{A}\Rpar{\xi,\frac{R_b}{R_a}}}{\log\frac{R_b}{R_a}-\log\Xi-\mathcal{B}\Rpar{\xi}}.\\
\label{eq:heuristic_f_approx}
\ffin
\Cref{eq:heuristic_f_approx} provides the leading order and the main
functional behavior for the condensed fraction of ions in the region
below saturation.
In the limit where $\log\Rpar{{R_b}/{R_a}}$ is {\it large} we obtain the dominant behavior as
\eemp
\frac{f-f_M}{f_M}\approx&\frac{\log\Xi-\mathcal{A}\Rpar{\xi,\frac{R_b}{R_a}}}{\log\frac{R_b}{R_a}}.
\label{eq:heuristic_f_func_form}
\ffin
It is noteworthy that we recover the same expression as in the 
empirical Eq.~(\ref{eq:f_empirical}).

\psfrag{AX}{$\log(R_b/R_a)$}
\psfrag{BY}{$(f-f_M)/f_M$}
\ImT{0.5}{art28_size_D_new_vfinal}{size_f2_w_f}{Condensed fraction under strong-coupling for $\xi=4$ and $5$ in a log-log plot where error bars are less than the tick size. The solid curve represents the numeric solution obtained for \cref{eq:heuristic_f} and the symbols are from Monte Carlo. Notice that the analytic solution predicts saturation near $\log\frac{R_b}{R_a}\approx10$ for $\Xi=10^3$ and $\log\frac{R_b}{R_a}\approx20$ for $\Xi=10^4$. {(Inset)} Numerical results for the condensed fraction for $\xi=4$ and the approximation from \cref{eq:heuristic_f_func_form}.}

The numerical results from Monte Carlo simulations are displayed in \cref{fig:size_f2_w_f,fig:size_f3_w_f}, and compared both 
to the analytic prediction solving numerically \cref{eq:heuristic_f} and to the asymptotic large box size expansion of \cref{eq:heuristic_f_func_form}. The agreement is good; 
we are indeed in the relevant regime of parameters where $\Xi \gg 1$, with the additional needle constraint $\xi^2 \ll \Xi$.
\emma{These results tell us that we should expect saturation below a critical box size and above a certain coupling. Finally, 
regardless of the box size, $f\to f_M$ when $\Xi<e^{\mathcal{A}}$. The quantity $\mathcal{A}$, as shown in \cref{fig:size_delta_behavior}, 
compares relatively well with the numerical estimation of $4.5$ reported in \cite{MTT13}.}

\psfrag{AX}{$\Xi$}
\psfrag{BY}{$\Rpar{\log(R_b/R_a)}(f-f_M)/f_M$}
\ImT{0.5}{art29_size_Xi_new_vfinal}{size_f3_w_f}{Condensed fraction under strong-coupling for $\log(R_b/R_a)=30$ in a log-linear plot. 
As in Fig \ref{fig:size_f2_w_f}, the symbols are for Monte Carlo data and the curves for the solution of \cref{eq:heuristic_f}.
The analytic result predicts saturation, visible only for $\log\Rpar{R_b/R_a}=30$ close to $\Xi\approx2\times10^4$. {(Inset)} Monte Carlo results for the condensed 
fraction compared the approximation from \cref{eq:heuristic_f_func_form}.}


\psfrag{AX1}{$\log(R_b/R_a)$} \psfrag{AX2}{$\xi$}
\psfrag{BY}{$\mathcal{A}\Rpar{\xi,\frac{R_b}{R_a}}$}
\ImT{0.4}{size_delta}{size_delta_behavior}{Plot of $\mathcal{A}$ as a
  function of $\log(R_b/R_a)$ and $\xi$. The value obtained in
  \cite{MTT13} corresponds to $\mathcal{A}\simeq4.5$. Also, we used
  the abbreviation $\Delta=\log(R_b/R_a).$}

\subsubsection{Strong coupling -- Thick cylinder limit}

In this subsection, we consider the strong coupling regime $\Xi\gg 1$,
but for a thick cylinder, i.e. $\xi/\sqrt{\Xi} \gg 1$. In this limit,
the radius $R_a$ of the cylinder is much larger than the lattice
constant $a$ of the Wigner crystal of counterions formed in the limit
$T=0$ in the surface of the cylinder. \emma{This is depicted in Fig. \ref{fig:phase_diag2}-c).}
The Wigner crystal is almost the
planar hexagonal lattice, with probably some defects to accommodate to
the curvature of the cylinder. The lattice spacing $a$ is given by~$a
= c R_a \sqrt{\Xi}/\xi$, with $c=(4\pi/\sqrt{3})^{1/2}$. Let us consider
the situation $R_b=\infty$ and focus on the contact density
at the charged cylinder. In the planar case ($R_a=\infty$), it is
$\widetilde{\rho}(R_a)=1$. We wish to estimate here the first correction
to this value due to the cylinder curvature.

Rescaling all lengths by $a$ in (\ref{eq:cont-cyl-ocp-Rbinfty}) gives
\begin{equation}
  \label{eq:rhoRathick1}
  \widetilde{\rho}(R_a) \,=\,  2 \left( \frac{\xi-1}{\xi}\right )^2
  -\frac{\xi-1}{2\xi c} \, \langle \widetilde{w}\rangle \, \frac{\sqrt{\Xi}}{\xi}
\end{equation}
with $\widetilde{w}=\sum_{i\in B} (\widetilde{r}_i^{\perp})^2/\widetilde{r}_i^3$,
where $\widetilde{\mathbf{r}}_i=\mathbf{r}_i/a$ are located at the
positions of the crystal arrangement.  The summation in $\widetilde w$ involves the bound ions,
but we are here in a limit where the fraction of bound ions is very close to unity
(from previous sections, we have that $1>f>f_M=1-1/\xi$, and $\xi$ has to be large,
meaning that $f\simeq 1$). We will therefore neglect the unbound ions in this analysis.
Since the surface is almost
planar (the curvature is measured by the ratio $a/R_a \ll 1$),
we can approximate $\widetilde{w}=S_1+S_2$ with
\begin{equation}
  \label{eq:S1def}
  S_1=\sum_{n} \sum_{j=-\infty}^{+\infty} \frac{n^2}{(n^2+3 j^2)^{3/2}}
\end{equation}
and
\begin{equation}
  \label{eq:S2def}
  S_2=\sum_{n} \sum_{j=-\infty}^{+\infty} 
  \frac{(n+1/2)^2}{((n+1/2)^2+3 (j+1/2)^2)^{3/2}}
\end{equation}
The sum over $n$ runs over all lattice points on the circumference of
the cylinder. That sum gives a leading contribution which is of order
$2\pi R_a /a$, the number of lattice points in the
circumference. Therefore, using the fact that
\begin{equation}
  \label{eq:s1limit}
  \lim_{n\to\infty}\sum_{j=-\infty}^{+\infty}
    \frac{n^2}{(n^2+3 j^2)^{3/2}}
    = \frac{2}{\sqrt{3}}
    \,,
\end{equation}
it is convenient to write
\begin{equation}
  S_1= \frac{2\pi R_a}{a} + \sum_{n=-\infty}^{+\infty}\left[ \sum_{j=-\infty}^{+\infty}
    \frac{n^2}{(n^2+3 j^2)^{3/2}}
    - \frac{2}{\sqrt{3}}\right]
\end{equation}
and a similar equation for $S_2$. Notice that the regularized sum
\begin{equation}
  \label{eq:S1reg}
  S_1^{*}=\sum_{n=-\infty}^{+\infty}\left[ \sum_{j=-\infty}^{+\infty}
    \frac{n^2}{(n^2+3 j^2)^{3/2}}
    - \frac{2}{\sqrt{3}}\right]
\end{equation}
is convergent, and can be numerically evaluated:
$S_1^{*}\simeq -0.80959$. The regularized version of the second sum is
$S_2^{*}\simeq -1.29712$. Putting all results together
\begin{equation}
  \label{eq:rhoRathick}
  \widetilde{\rho}(R_a)=
  \left(\frac{\xi-1}{\xi}\right)^2
  \left(
    1 -\frac{1}{\xi-1}
    -\frac{\xi}{\xi-1} \frac{S_1^{*}+S_2^{*}}{2 c} 
    \frac{\sqrt{\Xi}}{\xi}
  \right)
\end{equation}
Taking into account that $1/\xi$ corrections are sub-leading terms
compared to $\sqrt{\Xi}/{\xi}$ (since $\xi\gg\sqrt{\Xi}\gg1$), and the numerical values of the
lattice sums $S_1^{*}$ and $S_2^{*}$, we find
\begin{equation}
  \label{eq:rhoRathick-numeric}
  \widetilde{\rho}(R_a) \,=\, f_{M}^2\Rpar{1 + s 
    \frac{\sqrt{\Xi}}{\xi}} \, 
    \emma{\simeq \Rpar{1 + s 
    \frac{\sqrt{\Xi}}{\xi}}}
    \qquad \text{with\ } s\defeq-\frac{S_{1}^{*}+S_{2}^{*}}{2c}\simeq 0.391066
    \ .
\end{equation}
To leading order and for the sake of comparison with results in spherical geometry, this can be rewritten as 
\begin{equation}
 \widetilde{\rho}(R_a) = \frac{n(R_a)}{2\pi\ell_B \sigma_a^2 } \, \simeq\, 1 +
\sqrt{\frac{q}{\pi \sigma_a}} \, \frac{s}{\sqrt{2}} \, \mathcal{C} 
\label{eq:cylthickSC_bis},
\end{equation}
where $\mathcal{C}=1/R_a$ is the curvature of the colloid.

We have assumed here the lattice on the cylinder to be arranged such that
sites separated by a distance $a$ lie on the circumference of the
cylinder. One could also consider that the lattice is arranged so
that the sites on the circumference are separated by $\sqrt{3} a$,
that is, the previous lattice rotated by $\pi/2$. The individual sums
$S_1$ and $S_2$ change. For instance the regularized equivalent of
$S_1$ will be
\begin{equation}
  \label{eq:S1regbis}
  S_1^{**}=\sum_{n=-\infty}^{+\infty}\left[ \sum_{j=-\infty}^{+\infty}
    \frac{3n^2}{(3n^2+ j^2)^{3/2}}
    - 2 \right]  \,.
\end{equation}
Although the individual sums are different, their sum
$S_1^{*}+S_2^{*}$ is the same, yielding the final result
(\ref{eq:rhoRathick}) unchanged. To put this prediction to the test,
we plot in Fig. \ref{fig:contact_R_def_SC} the contact density 
in a way that clearly evidences the correction embedded in 
expression (\ref{eq:rhoRathick}). When $\xi>\sqrt{\Xi}$ (the thick cylinder range), the agreement
is noticeable. This analysis confirms the relevance of the parameter $\xi / \sqrt{\Xi}$
as ruling the strong coupling large $\Xi$ regime.

\psfrag{AX}{$\xi/\sqrt{\Xi}$}
\psfrag{BY}{$\widetilde{\rho}/f_{M}^2-1$}
\ImT{0.4}{contact_R_def}{contact_R_def_SC}{Rescaled contact density at $R_a$ $\Rpar{\widetilde{\rho}/f_M^2-1}$ as a function of $\xi/\sqrt{\Xi}$, on a log-log scale. 
The crossover from the needle into the thick limit appears approximately at $\xi/\sqrt{\Xi}\approx0.4$. The dashed curves represents the prediction for the thin (black -- \cref{eq:rhoRa} without the correction in $\zeta(3)$, valid on the left hand side) and thick (blue -- \cref{eq:rhoRathick}) cases.}

Figure \ref{fig:contact_R_def_SC} exemplifies the crossover between the thin and the thick cylinder limiting cases.  
A further illustration is provided by Fig. \ref{fig:contact_R_SC0}, showing the contact density. At small $\Xi$, it is not a surprise to see
the mean-field result hold. At large $\Xi$ ($10^2$ and $10^3$ on the figure), the behavior depends on the ratio
$\xi/\Xi^{1/2}$. If it is small, the thin cylinder formula applies (dotted line), and leads to a rescaled
contact density which increases with $\xi$, at fixed $\Xi$. On the other hand, for $\xi>\Xi^{1/2}$,
the thick cylinder phenomenology takes over and leads to a decrease of contact density. 
The maximum of the curve corresponds to the crossover between both regimes, that is again found
to take place at $\xi/\sqrt{\Xi}\approx0.4$. It appears that for all fixed $\Xi$, no matter how large, 
the limit of large $\xi$ invariably leads to $\widetilde \rho(R_a) = 1$. This was expected, since
this is nothing but the contact theorem for a planar interface, which holds for 
all values for $\Xi$. 

\psfrag{AX}{$\xi$}
\psfrag{BY}{$\widetilde{\rho}\vert_{r=R_a}$}
\ImTc{0.4}{contact_R0}{contact_R_SC0}{Contact density at ${r}={R_a}$ versus Manning parameter $\xi$ for $\Xi=0.1,\, 10^2$ and $10^3$; 
here $R_b/R_a=e^{300}$ and $N=300$. The symbols are for the Monte Carlo data. The dashed line shows the mean-field (Poisson-Boltzmann)
prediction \eqref{eq:rhocMFcyl}, very accurate for small $\Xi$.
The dotted curve displays the leading order form \cref{eq:rhoRa}, which does not depend on $\Xi$. The solid red and green curves represent the 'thick' prediction from \eqref{eq:rhoRathick-numeric} respectively for $\Xi=10^2$ and $10^3$. The solid arrowed horizontal lines represent 
the asymptotic infinite $\xi$ needle (top) and thick (bottom) limits. Note that \emma{at large enough $\Xi$, the planar limit (large $\xi$)
is approached from above, with contact densities always larger than unity, at variance with mean-field behavior}.}{}

To conclude this study of screening in cylindrical geometry (with 3D Coulombic interactions in $1/r$ between particles),
we have summarized in Fig. \ref{fig:phase_diag2} our main findings pertaining to the contact density. In all the present 
subsection, the system size $R_b/R_a$ can be considered as (exponentially) large, so that the results 
presented apply to an isolated charged macro-cylinder.



\section{Application II: Screening of a spherical macro-ion and effect of Coulombic coupling}
\label{sec:applII}

After having focused on the screening properties of cylindrical macro-ions, we  
will address the case of spheres in $d=3$. We start by the strong coupling limit, where mean-field
breaks down, and then turn to the weak coupling limit. It will be shown 
that the effect of curvature on the contact density is opposite in these two limiting cases.
This conclusion also holds for cylinders.

\subsection{Strong coupling}
\label{ssec:sphstrong}

The analysis of section \ref{sec:3d} provides a convenient starting point for discussing strong coupling effects
for spherical colloids, a topic of interest \cite{Shkl99,Levin02,WSC1,WSC2}. In particular,
Monte Carlo simulations have been performed \cite{LiLo00}, where the density of counterions 
in contact with a spherical macro-ion is reported at various couplings. The system studied 
is salt free, so that we resort to Eq. \eqref{eq:cont3docp}. The Coulombic coupling within an ensemble 
of colloidal spheres is measured again as 
$\Xi=2\pi \ell_B^2 q^3 |\sigma_a|$;
it may be enhanced by increasing the valency $q$ and we consider the often studied case $\sigma_b=0$, $\sigma_a\neq 0$.

In general, the mean Hamiltonian, which appears in \eqref{eq:cont3docp}, is 
not known explicitly, but a useful simplification occurs  
when $\Xi$ becomes large enough (roughly speaking, larger than 50).
Indeed, most counterions lie in the immediate vicinity of the colloids, and form
a strongly modulated two dimensional liquid, or crystal if $\Xi$ exceeds the 
crystallization threshold. In these conditions, a one component plasma picture 
may be invoked \emma{(with a Wigner hexagonal crystal of counter-ions in 
a neutralizing two-dimensional background)}, and an excellent approximation 
for the energy $\langle H \rangle $ is given by its ground state value $U$,
which reads \cite{rque150}
\begin{equation}
\beta U  \,=\, -\mathcal{M} \,\frac{\ell_B}{2 R_a} \, Q_a^{3/2} q^{1/2}
\label{eq:Uground}
\end{equation}
where $\mathcal{M} \simeq 1.10$ is some Madelung constant \cite{Levin02}.
In the above equation, curvature effects have been neglected, and the energy
expressed from its planar limit. This requires that the number of charges $Q_a=4\pi R_a^2 \sigma_a$
is somewhat larger than unity, a condition that is easily met in practice. 
Enforcing $\langle H \rangle = U$ in Eq. \eqref{eq:cont3docp},
we arrive at 
\begin{equation}
\frac{n(R_a)}{2\pi\ell_B \sigma_a^2 } \, \simeq\, 1 + \mathcal{M} \left(\frac{q}{Q_a} \right)^{1/2}
+ 6 \,\frac{R_a}{q Q_a \ell_B} \left[ \frac{P}{P_{id}} \,-1
\right]
\label{eq:int_sph}
\end{equation}
where it was remembered that $\beta P = n(R_b)$ is the pressure of the system,
and $P_{id}$ is the ideal gas reference pressure ($\beta P_{id} = Q_a \rho_c/q$,
where $\rho_c$ is the colloidal mean density). The contribution 
in $R_b^3 \,n(R_b)$, giving rise to the term in $P/P_{id}<1$ in \eqref{eq:int_sph} usually is negligible at large couplings, except at very large
colloidal concentrations, so that we have
\begin{equation}
\widetilde\rho(R_a)\,=\, \frac{n(R_a)}{2\pi\ell_B \sigma_a^2 } \, \simeq\, 1 + \mathcal{M} \left(\frac{q}{Q_a} \right)^{1/2}
- 6 \, \frac{R_a}{q Q_a \ell_B} .
\label{eq:sphereSC}
\end{equation}
To test this prediction, we consider that parameter set in Ref.~\cite{LiLo00}
that exhibits the largest coupling: $Q_a=60$, $q=3$, $R_a=2.2\,$nm \cite{rque151}.
Upon changing the density of colloids by a factor of 8, the ionic density at contact $n(R_a)$,
as given in Table II of Ref.~\cite{LiLo00}, is remarkably constant, between 4.9 and 5 $10^{-6}\,\hbox{nm}^{-3}$.
Making use of Eq. \eqref{eq:sphereSC} gives 5 $10^{-6}\,\hbox{nm}^{-3}$, in excellent agreement.
This correspond to an increase of the contact density, compared to the planar limit, by
an amount of 14\%. Indeed, in Eq. \eqref{eq:sphereSC}, the dominant correction 
on the right hand side is $\mathcal{M} \left(q/Q_a \right)^{1/2}$, and thus positive:
the effect of curvature is here to enhance ionic density at contact (we are in a limit
where the presence of the outer boundary at $r=R_b$ does not affect the profile at $r=R_a$).
A way to decrease curvature, at fixed surface charge $\sigma_a$, would be to increase 
$R_a$, and given that $Q_a \propto R_a^2$, Eq. \eqref{eq:sphereSC} yields the expected
unity on the right hand side in that limit. 
We stress here that the simulations in Ref \cite{LiLo00} were not performed
with the cell model restriction, but for a collection of 80 colloids (with thus
their $80\times 60/3 = 1600$ counterions). The agreement found not only assesses
the strong coupling approach \cite{rque152}, but also the cell viewpoint as such.

For comparison with results pertaining to the cylindrical geometry,
it is convenient to rewrite Eq. \eqref{eq:sphereSC} as 
\begin{equation}
\frac{n(R_a)}{2\pi\ell_B \sigma_a^2 } \, \simeq\, 1 +
\sqrt{\frac{q}{\pi \sigma_a}} \,\frac{1}{R_a} \left(\frac{\mathcal{M}}{2} -\frac{3}{\sqrt{2\Xi}}
\right)
\end{equation}
It is a strong coupling-weak curvature expansion, which reads, to dominant order in coupling
\begin{equation}
\frac{n(R_a)}{2\pi\ell_B \sigma_a^2 } \, \simeq\, 1 +
\sqrt{\frac{q}{\pi \sigma_a}} \, \frac{\mathcal{M}}{4} \,\mathcal{C} 
\label{eq:sphereSC_bis}.
\end{equation}
where $\mathcal{C} = 2/R_a$ is the curvature of the colloid.
Given that the quantity $s$ introduced in \eqref{eq:rhoRathick-numeric}
fulfills $s/\sqrt{2} = \mathcal{M}/4$, it appears that expression 
\eqref{eq:sphereSC_bis} coincides with \eqref{eq:cylthickSC_bis},
valid for weakly curved, strongly charged cylinders. We therefore surmise
that relation \eqref{eq:sphereSC_bis} is valid for all curved objects under strong coupling $\Xi$,
provided that the local radius of curvature is large compared 
to $\sqrt{q/\sigma_a}$, the lattice constant that would be formed at vanishing temperature.
The reason for the positive sign of the curvature correction is quite clear by considering  
{\em a contrario}\/ a negatively curved macroion where curvature brings counterions closer to
each other than in the planar case. The opposite happens here for positively 
curved objects, and curvature is thus conducive to ionic condensation 
onto the macro-ion.

\subsection{From strong to weak couplings}

We have seen that compared to a plate of similar surface charge, the ionic density at contact is enhanced due to curvature
for both spheres and cylinders. This has been shown explicitly in the strong coupling limit $\Xi\gg 1$, but there are hints 
from simulations \cite{LiLo00} that the opposite conclusion may hold in the mean-field (Poisson-Boltzmann) regime $\Xi \ll 1$.
For consistency with section \ref{ssec:sphstrong}, we take again $\sigma_b=0$ and the question is now to know the
sign for the quantity $n(R_a)- 2\pi\ell_B \sigma_a^2 -n(R_b)$, which identically vanishes
in the planar case. We start from the exact relation \eqref{eq:cont3docp}
\begin{equation}
n(R_a) -  2\pi\ell_B \sigma_a^2 -n(R_b) \,=\, 
-\frac{\beta \langle H\rangle}{4\pi R_a^3} + \left[\left(\frac{R_b}{R_a}\right)^3-1\right] n(R_b) 
\, - \, \frac{3|\sigma_a|}{q R_a}.
\label{eq:intsph2}
\end{equation}
that holds both under small or large couplings.
While one has in general $\langle H\rangle<0$, and $R_b>R_a$ by construction, the last term on the right hand side is negative,
so that no conclusion can be drawn at this stage concerning the sign of the
right hand side of the equality, even assuming $\Xi$ small. To proceed, another type of argument 
is necessary, and \eqref{eq:intsph2} must be relinquished. The idea is to invoke 
the Poisson-Boltzmann equation itself, fulfilled by the mean-field electric potential \cite{Levin02},
from which the ionic density follows:
\begin{equation}
\nabla^2 \phi \,= \, -4 \pi \ell_B \, \sum_\alpha q_\alpha n_\alpha^0 e^{-q_\alpha \phi}.
\label{eq:PB}
\end{equation}
This is the most general form for a mixture, where the ionic density for species
$\alpha$ is $n_\alpha^0 \exp(-q_\alpha \phi)$ and $\phi$ is dimensionless. We can treat here on equal footings the cylindrical
$d=2$
and the spherical $d=3$ cases, which only differ from the expression of the Laplacian $\nabla^2$.
We assume $\phi$ to be radially symmetric, multiply both sides of Eq. 
\eqref{eq:PB} by $\partial \phi/\partial r$, and integrate with respect to $r$.
This yields 
\begin{equation}
n(R_a) -  2\pi\ell_B \sigma_a^2 -n(R_b) \,=\, 
-(d-1) \int_{R_a}^{R_b} \frac{1}{r'} \left(\frac{\partial \phi}{\partial r'}\right)^2 \,dr' ,
\label{eq:PBcurv}
\end{equation}
which is therefore negative, as anticipated. 
Thus, at mean-field level, the effect of curvature is to decrease the contact ionic density
at $R_a$. This is the opposite scenario compared to that occurring under strong coupling (see section \ref{ssec:sphstrong}). 
We also note
that whenever $\sigma_b \neq 0$, relation \eqref{eq:PBcurv} holds provided
the left hand side is replaced by $n(R_a) -  2\pi\ell_B \sigma_a^2 -n(R_b) + 2\pi\ell_B \sigma_b^2$.
We stress again that the conclusion on the sign also holds in two dimensions,
and for $d=1$ (planar case), the right hand side of \eqref{eq:PBcurv} vanishes,
see the constraint \eqref{eq:contact_planar}, that is (remarkably) also valid within mean-field.
For $d=2$, it was already seen in Fig. \ref{fig:contact_R_SC0} that the $\Xi=0.1$ results 
were below unity, meaning that $n(R_a) <  2\pi\ell_B \sigma_a^2$ (we are there close to
the $R_b \to \infty$ limit for which $n(R_b)\to 0$, with furthermore $\sigma_b=0$).

A further comment concerning Poisson-Boltzmann theory is in order.
Within the mean-field premises, the internal energy of the system may be
expressed
\begin{equation}
\beta \langle H\rangle \,=\, \frac{1}{2} \int_{cell} \frac{1}{e} \, \rho_{tot}(\mathbf{r}) \phi(\mathbf{r})\,d\mathbf{r}
\end{equation}
where the integral, running over the entire cell, involves the total ionic density
$\rho_{tot}$, and includes also the charged boundaries at $R_a$ and $R_b$.
Inserting this relation in \cref{eq:intsph2}, we obtain a 'sum rule',
that holds at the level of Poisson-Boltzmann theory only, but that can be proven 
starting directly from \eqref{eq:PB} \cite{Ladislav}. This is another confirmation, in a limiting
case, for the validity of the expressions we have derived.


\section{Conclusion}

The exact contact relation 
\begin{equation}
n_a - 2 \pi \ell_B \sigma_a^2 - n_b + 2 \pi \ell_B \sigma_b^2 =0 
\label{eq:concl}
\end{equation}
does only hold in the planar case, when classical ions interacting by Coulomb
forces are confined in a slab of two parallel walls, bearing surface charge densities
$\sigma_a e $ and $\sigma_b e $. In itself, this relation is remarkable, for it does not depend 
on the strength of Coulombic coupling. It therefore equally applies to weakly, moderately, and strongly
coupled situations, and is therefore fulfilled, in particular, by the Poisson-Boltzmann mean-field theory
\cite{rque200}.
Our primary motivation was to investigate how it 
should be modified when dealing with curved interfaces. To this end, we considered
a cell model approach where a macro-ion of surface charge density $\sigma_a e $ is enclosed in a concentric 
confining cell of similar geometry, cylindrical, or spherical (with charge density $\sigma_b e $). 
New exact relations were derived. Quite expectedly, the
two-dimensional results (i.e. when charges interact through a $\log$ potential) 
are more explicit than their three dimensional counterpart. 
Our results provide a convenient starting point to 
discuss the strong coupling limit of the contact densities under study. 
In particular, it was shown (for $\sigma_b=0$, but with presumably no loss of
generality), that the l.h.s of Eq. \eqref{eq:concl} is positive for weak
curvatures, with an expression that is the same for both spherical and
cylindrical macro-ions, see Eq. \eqref{eq:sphereSC_bis} which embodies 
the exact strong-coupling correction to the planar case. For cylindrical macro-ions,
the situation of strong curvature was also worked out (referred to as the
needle, or thin cylinder limit). 
On the other hand, in the mean-field limit, that is when the Coulombic 
coupling measured by a parameter of the form $\ell_B^2 \sigma_a$ is small,
the quantity on the l.h.s.~of \cref{eq:concl} becomes negative,
a phenomenon that does not seem particularly intuitive.

We would like to thank Ladislav \v{S}amaj, Alexandre Pereira dos
Santos and Yan Levin for fruitful discussions. The authors acknowledge
support from ECOS-Nord/COLCIENCIAS-MEN-ICETEX. J.P.M. and
G.T. acknowledge partial financial support from Fondo de
Investigaciones, Facultad de Ciencias, Universidad de los Andes.

\appendix

\section{Derivation of Eq. \eqref{eq:cont-cyl}}
\label{app:h_cyl}

Let
\begin{eqnarray}
  W_1/e^2&=& \left\langle \sum_{i<j} q_{i}q_{j}
  \frac{{r_{ij}^{\perp}}^2}{r_{ij}^3} \right\rangle\\ &=& \frac{1}{2}
  \int \sum_{\alpha \gamma} q_{\alpha} q_{\gamma}\,
  n_{\alpha}n_{\gamma}(1+h_{\alpha\gamma}(\r_1^{\perp},\r_2^{\perp},z_1-z_2))
  \frac{{|\r_1-\r_2|^{\perp}}^2}{|\r_1-\r_2|^3}\, d\r_1d\r_2
\,.
\end{eqnarray}
We have
\begin{equation}
  W_1/e^2=\lim_{L\to\infty}\frac{L}{2}\int d^2\r_{1}^{\perp}d^2\r_{2}^{\perp}
  \int_{-L/2}^{L/2} dz \sum_{\alpha \gamma} q_{\alpha} q_{\gamma}\,
  n_{\alpha}n_{\gamma}(1+h_{\alpha\gamma}(\r_1^{\perp},\r_2^{\perp},z))
  \frac{{r_{12}^{\perp}}^2}{\left({r_{12}^{\perp}}^2+z^2\right)^{3/2}}
\end{equation}
Using
\begin{equation}
  \label{eq:useful-integral}
  \int_{-L/2}^{L/2} \frac{dz}{\left({r_{12}^{\perp}}^2+z^2\right)^{3/2}} =
  \frac{2L}{{r_{12}^{\perp}}^2\sqrt{{r_{12}^{\perp}}^2+L^2}} \to
  \frac{2}{{r_{12}^{\perp}}^2},\qquad (L\to\infty)
\end{equation}
we find
\begin{eqnarray}
  W_1/e^2&=&\frac{1}{L}( Q_a+Q_b)^2
  \nonumber\\
  &&+\frac{L}{2}
  \int d^2\r_{1}^{\perp}d^2\r_{2}^{\perp}
  \int_{-L/2}^{L/2} dz \sum_{\alpha \gamma} q_{\alpha} q_{\gamma}\,
  n_{\alpha}n_{\gamma}h_{\alpha\gamma}(\r_1^{\perp},\r_2^{\perp},z)
  \frac{{r_{12}^{\perp}}^2}{\left({r_{12}^{\perp}}^2+z^2\right)^{3/2}}
  \,.
  \nonumber
  \\
\end{eqnarray}
The virial average is then
\begin{equation}
  \langle W \rangle /e^2 = \frac{1}{L} (Q_{b}^{2}-Q_{a}^2)
  +\frac{L}{2}
  \int d^2\r_{1}^{\perp}d^2\r_{2}^{\perp}
  \int_{-L/2}^{L/2} dz \sum_{\alpha \gamma} q_{\alpha} q_{\gamma}\,
  n_{\alpha}n_{\gamma}h_{\alpha\gamma}(\r_1^{\perp},\r_2^{\perp},z)
  \frac{{r_{12}^{\perp}}^2}{\left({r_{12}^{\perp}}^2+z^2\right)^{3/2}}
\end{equation}
Replacing in \cref{eq:cont1}, we find \cref{eq:cont-cyl}.

\section{Screening of cylindrical macroions : Condensed fraction of ions}
\label{APP_towards_f}


Our starting point is the contact balance equation \cref{eq:contact_balEQ}.
We then introduce the potential of mean force $\Phi$ such that 
\eemp
\widetilde \rho(r)\, \propto\, e^{-\beta\Phi(r)} .
\ffin
As such, the potential carries contributions from the charged rod, the bound ($\mathcal{B}$) and unbound ($\mathcal{U}$) charges as,
\eemp
\beta\Phi(r)&=2\xi\log\frac{r}{R_a}+\beta\Phi_\mathcal{B}(r)+\beta\Phi_\mathcal{U}(r),
\label{eq:betaPhi_whole}
\ffin
with $2\xi\log r$ the energy due to the cylinder, and $\Phi_\mathcal{B}$ and $\Phi_\mathcal{U}$ respectively to the ions.
With these notations:
\eemp
\Rpar{\frac{R_b}{R_a}}^2\frac{\widetilde{\rho}(R_b)}{\widetilde{\rho}(R_a)}=&\exp\(2\log\frac{R_b}{R_a}+\beta\Phi(R_a)-\beta\Phi(R_b)\)\\
=&\exp\(-2(\xi-1)\log\frac{R_b}{R_a}+\beta(\Phi_\mathcal{B}(R_a)-\Phi_\mathcal{B}(R_b))+\beta(\Phi_\mathcal{U}(R_a)-\Phi_\mathcal{U}(R_b))\) .
\label{eq:ratiorho}
\ffin
The difficulty is now to obtain relevant expressions for the potentials
$\Phi_\mathcal{B}$ and $\Phi_\mathcal{U}$. This is the purpose of the following calculations.


\emma{The contribution $\Phi_\mathcal{U}$ stems from the very dilute cloud of of unbound ions, far from the charged cylinder,
and is of mean-field type. It can thus be obtained analytically, see below. On the other hand, the contribution from bound
ions is more difficult to estimate, and we will resort to a near-field expansion, when Coulombic coupling is large}.
Considering a perfectly formed one-dimensional Wigner crystal at $r=R_a$ (inner cylinder), the approach consists in taking the single particle variant energy 
formulation \cite{MTT13};
thus, we may write for the bound contribution,
\eemp
\beta\Phi_\mathcal{B}(r)=2\xi f\, G\Rpar{\frac{r}{a^\prime}}=2\xi f\, G\Rpar{\frac{\xi^2f}{\Xi}\frac{r}{R_a}},
\ffin
with $a^\prime$ the lattice spacing parameter of the crystal \emma{[see Fig \ref{fig:phase_diag2}a)]}, 
related to the parameters through \cite{MTT13} $R_a/a^\prime=\xi^2f/\Xi$. Here, 
$G(x)$ is defined as,
\eemp
G(x)\defeq&\sum_{j=1}^{\infty}\(\frac{1}{\sqrt{j^2+x^2}}-\frac{1}{j}\).
\label{eq:g_of_x}
\ffin
An approximate evaluation can be obtained through direct integration of
\cref{eq:g_of_x} using Euler--Maclaurin's formula. The result for $G(x)$ is,
\eemp
G(x)\simeq&-\log\frac{1+\sqrt{1+x^2}}{2}-\gamma\(1-\frac{1}{\sqrt{1+x^2}}\).
\ffin
Note that the small $x$ behavior for $G(x)$, which is
$G(x)\approx-\zeta(3)x^2/2$, is responsible for the
  corrections to the profile to leading order at $R_a$
  \cite{WSC1,MTT13}. For large $x$
  its behavior is given by $G(x)\approx-\log x +\log 2 - \gamma$.

The contribution from the unbound ions can be obtained under the assumption that the behavior of such ions is mean field
like. This population is subjected to the dressed potential of the inner cylinder, with an effective charge
$\xi_{\text{eff}}=\xi(1-f)$ that is smaller than unity as we have seen from previous results \cite{MTT13}. Such evaluation is possible
directly from the mean field density $n_{MF}$, through
\eemp
e^{\beta(\Phi_\mathcal{U}(R_a)-\Phi_\mathcal{U}(R_b))}=&\frac{n_{MF}(R_a)}{n_{MF}(R_b)}-2\xi_{\text{eff}}\log\frac{R_b}{R_a}.
\ffin
Notice we have substracted the contribution due to the effective cylinder as it is already
accounted for in \eqref{eq:betaPhi_whole}. Using Eq. (4) from \cite{MTT13} we obtain,
\eemp
\beta(\Phi_\mathcal{U}(R_a)-\Phi_\mathcal{U}(R_b))=2(1-\xi_{\eff})\log\frac{R_b}{R_a}+\begin{cases}
\log\Bpar{\frac{(\xi_{\eff}-1) ^2-\alpha^2}{1-\alpha^2}},&\xi_{\eff}\leq\xi_c \\
\log\Bpar{\frac{(\xi_{\eff}-1) ^2+\alpha^2}{1+\alpha^2}},&\xi_{\eff}\geq\xi_c \\
\end{cases}\,  .
\label{eq:deltaU_MF}
\ffin
Here $\xi_c$ is the Fuoss critical parameter defined as,
\emp
\xi_c=\frac{\log\frac{R_b}{R_a}}{1+\log\frac{R_b}{R_a}}\,  , 
\fin
and $\alpha$ is defined through a trascendental equation depending on $\xi_{\eff}$ and $\log(R_b/R_a)$.

Our interest in the computation of the condensed fraction of ions lies in the dilute regime where 
$f$ tends to $f_M$, or equivalently when $\log\Rpar{R_b/R_a}$ is \emph{large}. 
In this range, we may take $\xi_{\eff}\sim\xi_c$
(close to unity) where $\alpha=0$. Hence,
\emp
\beta(\Phi_\mathcal{U}(R_a)-\Phi_\mathcal{U}(R_b))\approx\, -2 \Rpar{\log\Bpar{\log\frac{R_b}{R_a}}-1}+\mathcal{O}\Bpar{\frac{1}{\log(R_b/R_a)}}.
\fin
Gathering results, 
\eemp
\beta\Phi_\mathcal{B}(R_b)-\beta\Phi_\mathcal{B}(R_a)\simeq& -2f\xi\log\frac{R_b}{R_a}+2f\xi\Bpar{\log\Xi-2\log\xi-\log f+\log 2-\gamma},
\label{eq:contact_contrib_B}
\ffin
\eemp
\beta\Phi_\mathcal{U}(R_b)-\beta\Phi_\mathcal{U}(R_a)\simeq&\, -2 \Rpar{\log\Bpar{\log\frac{R_b}{R_a}}-1},
\label{eq:contact_contrib_U}
\ffin
where $\gamma$ is Euler-Mascheroni constant. \Cref{eq:ratiorho} then
becomes
\begin{equation}
  \label{eq:ratio-altern}
  \Rpar{\frac{R_b}{R_a}}^2\frac{\widetilde{\rho}(R_b)}{\widetilde{\rho}(R_a)}
  =\exp\left(2(1-\xi+f\xi)\log\frac{R_b}{R_a} 
  +2\log\Bpar{\log\frac{R_b}{R_a}}-2
  -2f\xi\left(\log\Xi-2\log\xi-\log f+\log 2-\gamma\right)
  \right)
\end{equation}
This last equation, together with the contact balance equation
\eqref{eq:contact_balEQ}, gives the fundamental relationship for
the fraction $f$, \cref{eq:heuristic_f} in the main text.


\begin{thebibliography}{99}
\bibitem{rque10}
Incidentally, it is interesting 
to note that the celebrated Poisson-Boltzmann theory does fulfill the 
contact theorem, whereas other approaches, aiming at improving upon it, do not
\cite{FrLe12}.
\bibitem{rque5}
Note that this relation does not depend on the particular valencies nor charges
of the microscopic species.
\bibitem{rque19}
When $\sigma_b\neq 0$, the notion of pressure should be considered with 
caution, as discussed for one-component plasmas in Ref. \cite{Cho80}.
\bibitem{rque20}
When there is a symmetry mismatch, such as with a charged disc
in a spherical cell, or a charged cylinder of finite size,
the pressure takes a more complicated form \cite{Antypov06,TrHa97,Raul00}.
\bibitem{rque75}
Formally, the limit can be taken considering that $\beta e^2\sigma_a^2$ is fixed,
and likewise for $\sigma_b$. Otherwise, the limit $\beta e^2 \to 0$ is trivial,
with uniform profiles, as would occur say for diverging temperature.
\bibitem{rque50}
The Manning parameter for the two dimensional systems is defined as
$
\xi=\beta\Vpar{Q_a q}/2,
$
and is related to the coupling constant $\Gamma$ through neutrality as $\xi=N\Gamma/2$.
\bibitem{rque71}
This definition of the number of condensed ions $N_c$ agrees
with~\cite[Eq.~(19)]{BO05} if $2/\Gamma$ is not an integer. It differs
by one when $2/\Gamma$ is an integer. In the case when $2/\Gamma$ is
an integer, one counterion is ``floating'' around and a precise
definition of condensed counterions (such as the one proposed above)
should be provided to properly compute $N_c$ and decide if the
``floating'' counterion is included or not in the condensed fraction.
\bibitem{rque150}
Note that both sides of Eq. \eqref{eq:Uground} depend 
on inverse temperature, so that $U$ is effectively independent on 
temperature, as it should for a ground state feature.
\bibitem{rque151}
The radius of the colloids in Ref. \cite{LiLo00} is 2 nm, but it should be augmented by the
ionic radius (0.2 nm) to get a meaningful comparison.
\bibitem{rque152}
As a further test, we may compare the prediction of Eq. \eqref{eq:Uground} to the values 
reported in Ref. \cite{LiLo00}. Attention should be paid to the fact that the internal energy 
reported in \cite{LiLo00} excludes the self term and should thus be compared
to $\beta U$ as following from \eqref{eq:Uground} minus $\ell_B Q_a^2/(2 R_a)$. 
For $q=3$, we find an agreement better than $2\%$, which can furthermore be improved
by considering a more accurate expression for $U$, including sub-leading corrections
in the coupling parameter, as proposed for instance in \cite{Totsuji78}.
\end{thebibliography}
\end{document}